\documentclass{elsart}
\usepackage{natbib,amsmath,amssymb,latexsym,url,graphicx}
\usepackage{graphics}
\usepackage{psfig}
\usepackage{color}

\def\deg{{\rm deg}}

\def\qed{\hfill  \framebox(5,5){}}

\def\Res{{\rm Res}}

\newtheorem{theorem}{{\bf Theorem}}
\newtheorem{remark}{{\bf Remark}}
\newtheorem{definition}[theorem]{{\bf Definition}}
\newtheorem{corollary}[theorem]{{\bf Corollary}}
\newtheorem{proposition}[theorem]{{\bf Proposition}}
\newtheorem{lemma}[theorem]{{\bf Lemma}}

\begin{document}

\begin{frontmatter}



\title{Topology of 2D and 3D Rational Curves}


\author[a]{Juan Gerardo Alc\'azar\thanksref{proy}},
\ead{juange.alcazar@uah.es}
\author[b]{Gema Mar\'{\i}a D\'{\i}az-Toca\thanksref{proy}},
\ead{gemadiaz@um.es}

\address[a]{Departamento de Matem\'aticas, Universidad de Alcal\'a,
E-28871 Madrid, Spain}
\address[b]{Departamento de Matem\'atica Aplicada, Universidad de
Murcia,  30100 Murcia, Spain}

\thanks[proy]{Supported by the Spanish `` Ministerio de
Ciencia e Innovacion" under the Project MTM2008-04699-C03-01 }


\begin{abstract}
In this paper we present algorithms for computing the topology of planar and space rational
curves defined by a parametrization. The algorithms given here work directly with the
parametrization of the curve, and do not require to compute or use the implicit equation of the curve (in the case of planar curves) or
of any projection (in the case of space curves). Moreover, these algorithms
have been implemented in Maple; the examples considered and the timings obtained show good performance
skills.
\end{abstract}
\end{frontmatter}

\section{Introduction}\label{section-introduction}

The topology of planar algebraic curves, implicitly given, is a
well-studied problem (see \cite{Arnon}, \cite{gianni},
\cite{LaloCompl}, \cite{Lalo}, \cite{Hong}, and the more recent
works \cite{Eigen}, \cite{seidel}, among others); more recently,
the problem for space algebraic curves has also received certain
attention (see \cite{JG-Sendra}, \cite{Diat}, \cite{ElKa}). In all
these works it is assumed that the curve is given by means of
implicit equations, and the  considered algorithms deal with the
curves in this form. However, in this paper we address the
problem, apparently not discussed up to now, of computing the
topology of a rational curve (i.e. constructing a planar or space
graph describing the shape of the curve) starting directly from
its parametrization, without computing or making use of the
implicit equation of the curve.  This question may be of special
interest in the field of computer-aided geometric design (CAGD), where many of the curves used
are rational and even directly provided in parametric
form (e.g. Bezier curves, B-splines, NURBS).

Perhaps the reason for the absence of previous studies in this
direction is the common belief that if the parametric equations of
a curve are available, the curve is easy to visualize. This is
essentially true, but if the goal is to get a global idea of how
the curve is like, then there are still some difficulties. On the
one hand, one should previously compute a parameter interval such
that the plotting of the curve over the interval shows the main
features of the curve; this includes handling the case of possible
missing points/branches (which happens if some point of the curve
is generated when the parameter of the curve tends to infinity,
see for example \cite{Andradas}). On the other hand, the plotting,
as pointed out by Gonzalez-Vega and Necula in the introduction to
\cite{Lalo}, may not always provide a clear idea of the topology
of the curve, and hence auxiliary tools for describing the shape
of the curve may be of help.

In this paper we address both planar and space rational curves. As in other topology algorithms, we require the
input curves to satisfy certain conditions that can be achieved
with generality. In the planar case it is
required that the curve has neither vertical asymptotes nor
vertical components, and that the parametrization is {\it proper}
(see Section \ref{sec-background}). Initially,
the algorithm works in a similar way to existing
algorithms, i.e. first one computes the critical
points and the points of the curve lying on the lines $x=\alpha_i$
containing some critical point, and then one
appropriately connects these points. However, the connection phase
is carried out not in the usual way, but taking
advantage of the fact that a  parametrization
 is available (see Theorem \ref{th-connect} in
Section \ref{plane-case}). More precisely, the algorithm computes how the {\it parameter} values are
connected; so, two points are joined whenever the
algorithm detects that the parameter values giving rise to them
need to be connected. In particular, and unlike many classical
algorithms, this strategy does not require the curves to be in generic position (as defined in \cite{Lalo}).


The method is specially profitable in the case of space rational
curves. Existing implicit algorithms compute the topology of the
curve by projecting it onto a plane (the $xy$-plane, in our case),
and then lifting to space the topology of this projection. In
\cite{JG-Sendra}, this lifting phase is carried out in general by
using a second, auxiliary projection; however, in \cite{Diat},
\cite{ElKa} no auxiliary projection is needed. In any case, the
lifting of the singularities of the projection is a delicate
operation. In our case, we use a similar strategy for 3D curves.
However, here the lifting operation (which is performed without
auxiliary projections) presents no difficulties since the space
points are identified by the parameter values giving rise to them
(previously computed when addressing the projection). In the case
of 3D curves, our requirements are: (i) the curve has no
asymptotes or components normal to the $xy$-plane; (ii) the
projection onto the $xy$-plane fulfills the requirements of the 2D
algorithm.

We have implemented the algorithms in Maple 13; outputs and
timings of several examples are given in Section \ref{Examp}. In
our implementation we give to the user the option of computing
isolated points of the curve or not. The reason for this is that
isolated points correspond to points generated by complex,
non-real, values of the parameter, and therefore they may not be
of interest for certain users; moreover, the number of isolated
points is certified by means of Hermite's method (see \cite{cox})
and therefore it may be time-consuming.

The structure of the paper is the following. In Section
\ref{sec-background} we provide the necessary background on
rational curves; hence, notions like properness, normality,
critical and singular points are reviewed here, jointly with
related results. In Section \ref{plane-case} we provide the
algorithm for the 2D case. In Section \ref{sec-space}, the
algorithm for the 3D case is given. Finally, in Section
\ref{Examp} we describe some details of the implementation, and we
provide the outputs and timings of different examples in 2D and
3D. The parametrizations used in the examples
are given in Appendix I and Appendix II.

\section{Background on Rational Curves}\label{sec-background}

In this section we briefly recall the background on affine rational curves
that we need in order to develop our results. So, in the sequel
we will consider an affine rational
curve ${\mathcal C}$ defined by a rational parametrization
\[\varphi(t)=\left(x_1(t),x_2(t),\ldots,x_n(t)\right)=\left(\displaystyle{\frac{p_1(t)}{q_1(t)}},\displaystyle{\frac{p_2(t)}{q_2(t)}},\ldots,\displaystyle{\frac{p_n(t)}{q_n(t)}}\right)\]
where $\gcd(p_1,q_1)=\gcd(p_2,q_2)=\cdots=\gcd(p_n,q_n)=1$ and $p_i(t),q_i(t)\in {\Bbb Z}[t]$ for all $i=1,\ldots,n$. In our case $n=2$ or $n=3$; so, we will usually write $x,y,z$ instead of $x_1,x_2,x_3$. Moreover, since the parametrization is assumed to be real, we have that ${\mathcal C}$ is a real curve (i.e. that it consists of infinitely many real points), although for theoretical reasons when necessary we will see the curve embedded in ${\Bbb C}^n$. Nevertheless, our goal will always be the description of the shape of its real part.



A point $P_0\in {\Bbb R}^n$ is {\it reached} by the
parametrization $\varphi(t)$ if there exists $t_0\in {\Bbb C}$
such that $\varphi(t_0)=P_0$; in this case, we will also say that
$t_0$ generates $P_0$. Notice that the value of the parameter
generating a real point may be either real or complex, and that
there may be points generated by several (real or complex) values
of the parameter. In this sense, we will say that the
parametrization $\varphi(t)$ is {\sf proper} if almost all points
of ${\mathcal C}$ are reached by just one value of the parameter
$t$, i.e. if $\varphi(t)$ is injective for almost all the points
of ${\mathcal C}$. So, if $\varphi(t)$ is proper then there are
just finitely many points of ${\mathcal C}$ generated by several
different values of the parameter, corresponding to the {\it
self-intersections} of the curve. In order to check whether
$\varphi(t)$ is proper, we will use the following criterion. Let
\[
\begin{array}{c}
\tilde{G}_1(t,s)=p_1(t)q_1(s)-p_1(s)q_1(t)\\
\tilde{G}_2(t,s)=p_2(t)q_2(s)-p_2(s)q_2(t)\\
\vdots \\
\tilde{G}_n(t,s)=p_n(t)q_n(s)-p_n(s)q_n(t)\\
\tilde{G}(t,s)=\gcd(\tilde{G}_1,\tilde{G}_2,\ldots,\tilde{G}_n)
\end{array}
\]
Then, the following theorem, directly deducible from Proposition 7
in \cite{Rubio} (see also Theorem 4.30 in \cite{SWPD}, for the
planar case), holds.

\begin{theorem} \label{th-charact-proper}
The parametrization $\varphi(t)$ is proper iff $\tilde{G}(t,s)=t-s$.
\end{theorem}

On the other hand, we will say that $\varphi(t)$ is {\sf normal} if
every point in ${\mathcal C}$ is reached by at least one value of
the parameter, i.e. if
$\varphi({\Bbb C})={\mathcal C}$. If $\varphi(t)$ is
not normal, then (see Proposition 4.2 in \cite{Andradas})
there is just one point of ${\mathcal C}$ non-reached by the parametrization,
namely the point \[P_{\infty}=\mbox{lim}_{t\to \pm \infty}\varphi(t)\]Notice that $P_{\infty}$ exists
if and only if
$\deg(p_i)\leq \deg(q_i)$ for all $i\in \{1,2,\ldots,n\}$. Furthermore, if $P_{\infty}$ exists,
it may still be reached by some (real or complex) value of the parameter. If we denote
$P_{\infty}=(a_1,a_2,\ldots,a_n)$, $P_{\infty}$ is reached iff \[\deg\left(\gcd(a_1q_1(t)-p_1(t),a_2q_2(t)-p_2(t),\ldots,a_nq_n(t)-p_n(t))\right)\geq 1\]
Also, observe that if $P_{\infty}$ exists then it is obtained as the limit of a
sequence of real points of ${\mathcal C}$, and therefore it cannot be isolated.
 Hence, if $P_{\infty}$ is reached by some value $t_a$ then it is a self-intersection
 of the curve, because it is a crossing of two branches of the curve, one corresponding to $t\to \pm\infty$ and the
 other corresponding to $t_a$. On the other hand,
 if $P_{\infty}$ exists but it is not reached, one can
 reparametrize the curve so that it is reached (see Theorem 7.30
 in \cite{SWPD}). However, reparametrizations may complicate the
 equations of the curve, or bring other difficulties, like
 improperness or the introduction of algebraic numbers. Hence, in our case whenever we meet this
 phenomenon, we will understand that this reparametrization has
 not been performed.

If every point of ${\mathcal C}$ is reachable via $\varphi(t)$ by real values of the parameter
 one says that $\varphi(t)$ is ${\Bbb R}$-{\sf normal}. We refer to \cite{Andradas}, \cite{SWPD} for a
 thorough study of this phenomenon. If $\varphi(t)$ is not ${\Bbb R}$-{\sf normal}, then there exist real
 points $P\in {\mathcal C}$ reachable only by complex values of the parameter. Moreover, the following
 result (see  Proposition 4.2 in \cite{Andradas}) clarifies the nature of these points.

\begin{proposition}\label{R-normal}
Let $\varphi(t)$ be a proper parametrization of ${\mathcal C}$.
Then $P\neq P_{\infty}$, $P\in {\mathcal C}\cap {\Bbb R}^n$ is non-reached by any real value of the
parameter if and only if it
is a real isolated point of ${\mathcal C}$.
\end{proposition}



\subsection{Critical Points of Planar Rational Curves}\label{add-back}

In the rest of the section we assume that $n=2$, i.e.
that ${\mathcal C}\subset {\Bbb R}^2$ is a real rational curve
 parametrized by $\varphi(t)=(x(t),y(t))$. Let $f\in {\Bbb R}[x,y]$
be its implicit equation; then we have the following classical definitions:

\begin{definition} \label{def-crit-points}
A point $P\in {\mathcal C}$ is  called: (a) {\sf a critical point}
if $f(P)=\frac{\partial f}{\partial y}(P)=0$; (b) {\sf a singular
point}, if it is critical and $\frac{\partial f}{\partial
x}(P)=0$; (c) {\sf a ramification point} if it is critical, but
non-singular; (d) {\sf a  regular point} if it is not critical.
\end{definition}

One may easily see that ramification points
correspond to those points satisfying that $x'(t)=0$ but $y'(t)\neq 0$, and that
singular points correspond to either points where $x'(t)=y'(t)=0$, or to
self-intersections of the curve. Singularities of a rational parametrization
can be computed directly from the parametrization, without converting to
implicit form. More precisely, the following result holds (see Theorem 10 and Theorem 11 in \cite{Sonia}).
Here, we denote $G_1=\tilde{G}_1/\tilde{G}$, $G_2=\tilde{G}_2/\tilde{G}$, and we write
$M(t)=\Res_s(G_1,G_2)$.

\begin{theorem} \label{th-sing-param}
Let $\varphi(t)$ be a parametrization of ${\mathcal C}$, and let $P_0\in {\mathcal C}$ be an affine singularity
of ${\mathcal C}$, reacheable by some value $t_0\in {\Bbb C}$ of the parameter. Then, $M(t_0)=0$.
\end{theorem}

\begin{remark} \label{inf-sing}
If $P_{\infty}$ is reached by some $t_0\in {\Bbb C}$ (in that case it is a
self-intersection of
the curve, and therefore a
singularity,
as we observed before), then $t_0$ must be a root of $M(t)$
(see Theorem 10 in
\cite{Sonia}).
\end{remark}

Whenever $\varphi(t)$ is proper, one may
deduce that $M(t)$ is not identically $0$; therefore, in that situation $M(t)$ has finitely many roots and from Theorem \ref{th-sing-param}, the $t$-values generating reachable singularities are among these roots. Now let us denote the numerator of $x'(t)$ by $N(t)$, and let us write the square-free part of $M(t)\cdot N(t)$ as $\tilde{m}(t)$; also, let $\tilde{q}(t)=\mbox{lcm}(q_1,q_2)$, and let $m(t)=\tilde{m}(t)/\gcd(\tilde{m}(t),\tilde{q}(t))$. Then, the following corollary on the real critical points of ${\mathcal C}$
can be deduced.

\begin{corollary} \label{gen-critical}
The real critical points of ${\mathcal C}$ are included in the (finite) set consisting of: (i) $P_{\infty}$ (if it exists); (ii) the real points generated by (real or complex) roots of $m(t)$.
\end{corollary}

\section{Computation of the Graph Associated with a Planar Curve} \label{plane-case}

Let ${\mathcal C}\subset {\Bbb R}^2$ be a planar algebraic curve,
parametrized by
\[\varphi(t)=(x(t),y(t))=\left(\displaystyle{\frac{p_1(t)}{q_1(t)}},\displaystyle{\frac{p_2(t)}{q_2(t)}}\right), \,\,  \gcd(p_1(t), q_1(t))=\gcd(p_2(t), q_2(t))=1 \]

In this section we address the problem of algorithmically
computing a graph ${\mathcal G}$ homeomorphic to the curve
${\mathcal C}$. In order to do so, we will follow the usual
strategy widely used in the implicit case (see \cite{Eigen},
\cite{Lalo}, \cite{Hong}, \cite{seidel}):
\begin{itemize}
\item [(1)] Compute the critical points of ${\mathcal C}$
(see Definition \ref{def-crit-points} in Subsection \ref{add-back}). Let $a_1<\cdots<a_m$ be the $x$-coordinates of
the critical points of ${\mathcal C}$; also, let $a_0=-\infty$, $a_{m+1}=+\infty$.
 \item [(2)] Compute the points of ${\mathcal C}$ lying on the
 vertical lines $x=a_i$, $i=1,\ldots,m$ passing through the critical points;
 we will refer to these lines as {\sf critical lines}.
 \item [(3)] For $i=1,\ldots,m-1$, compute the points of ${\mathcal C}$ lying on the vertical line $x=(a_i+a_{i+1})/2$; similarly for $x=a_1-1$, $x=a_m+1$. We will refer to these lines as ``non-critical" lines.
 \item [(4)] Connect, by means of segments, the points of ${\mathcal C}$ lying on each non-critical line, with the points in the critical lines immediately
 on its right and on its left, respectively.
     \end{itemize}

In our case, we will take advantage of the fact that a parametrization of the curve is available; this will be specially useful in order to carry out step (4). Moreover, in order to apply the method presented in this section, we need that certain hypotheses are fulfilled by ${\mathcal C}$. These hypotheses are introduced in Subsection \ref{subsec-hyp}. Then, in Subsection \ref{vertices} and Subsection \ref{edges} we show how to compute the vertices and edges, respectively, of the planar graph. Finally, in the last subsection we provide the full algorithm. The reader may find
several examples of the output of this algorithm in Section \ref{Examp}.

\subsection{Hypotheses}\label{subsec-hyp}

In the rest of the section, we assume that the following hypotheses are fulfilled:

\begin{itemize}
\item [(i)] $\varphi(t)$ is proper.
\item [(ii)] ${\mathcal C}$ has no vertical asymptotes; in particular, it is not a vertical line.
\end{itemize}

The first hypothesis guarantees that ${\mathcal C}$ is traced just
once when following the parametrization $\varphi(t)$. In order to
check this hypothesis, Theorem
\ref{th-charact-proper} can be applied. Moreover, if this hypothesis does not
hold, one can always reparametrize the curve (see Chapter 6.1 in
\cite{SWPD}) so that it is fulfilled. In order to check the second
hypothesis, one can use the following result, which is easy to prove.

\begin{lemma} \label{asymp-plane}
${\mathcal C}$ has a vertical asymptote iff one of the following
conditions occurs: (a) $q_2(t)$ has some real root which is not a
real root of $q_1(t)$; (b) $\deg(p_2)>\deg(q_2)$ but
$\deg(p_1)\leq \deg(q_1)$.
\end{lemma}

If ${\mathcal C}$ has some vertical asymptote, one proceeds in the following way:

\begin{itemize}
\item If  ${\mathcal C}$
has no horizontal asymptotes (which can be checked by
appropriately adapting Lemma \ref{asymp-plane}), then by interchanging
the axes $x$ and $y$ the condition is fulfilled. Notice that this is an affine
transformation, which therefore does not change the topology of the curve.
\item If ${\mathcal C}$ has also horizontal asymptotes, then
almost all changes of
coordinates of the type $\{x=X+\mu Y,y=Y\}$, with $\mu\in {\Bbb
Q}$, set the curve properly (see Proposition 3.2 in
\cite{LaloCompl}). Observe that if $\varphi(t)$ is proper, the curve ${\mathcal
C}_{\mu}$ obtained by applying such a transformation is properly
parametrized by $\varphi_\mu(t)=(x(t)-\mu y(t),y(t))$.
\end{itemize}

\subsection{Vertices of the Graph.}\label{vertices}

The notable points of ${\mathcal C}$ are the real critical points.
Now from Corollary \ref{gen-critical}, we have that these are
among the following points:

\begin{itemize}
\item [(i)] $P_{\infty}$ (if it exists).
\item [(ii)] The points of ${\mathcal C}$ generated (via $\varphi(t)$) by the real roots of the polynomial $m(t)$ in Corollary \ref{gen-critical}.
    \item [(iii)] The real points of ${\mathcal C}$ generated (via $\varphi(t)$) by complex roots of $m(t)$.
    \end{itemize}

    The computation of $P_{\infty}$ is described in Section
    \ref{sec-background}. Moreover, once the real roots of
    $m(t)$ are computed, the points in (ii) are obtained by
    evaluating $x(t),y(t)$ at these roots. Now we consider as vertices of the graph ${\mathcal
    G}$ not only these points, but
    also the points of ${\mathcal C}$ lying on the vertical lines
    containing the points in (i) and (ii). In order to compute these points, we recall
    the definition of the polynomials
$\tilde{G}_1,\tilde{G}_2,\tilde{G}$, introduced in Section
\ref{sec-background}, and we consider the polynomials
\[G_1(t,s)=\displaystyle{\frac{\tilde{G}_1(t,s)}{\tilde{G}(t,s)}},
\mbox{
}G_2(t,s)=\displaystyle{\frac{\tilde{G}_2(t,s)}{\tilde{G}(t,s)}}.\]
Then, given a point $P_r=(x_r,y_r)=\varphi(t_r)$, $t_r\in {\Bbb
R}$, the real roots of $G_1(t,t_r)$ provide the $t$-values of the
points lying in the line $x=x_r$; then, the coordinates of those
points can be obtained by evaluating $x(t),y(t)$ at these
$t$-values. Observe that we get not only the coordinates, but also
the $t$-values generating the points, via $\varphi(t)$. This is
important for the connection phase.

So, let us consider the points in (iii). If a point in (iii) is
also generated by a real value of the parameter, then it will have
already been computed as a point in (ii). So, if this is not the
case, by Proposition \ref{R-normal} it is an isolated point. Now
these points might
    be computed by seeking complex
    roots of $m(t)$ giving rise (when evaluating $x(t),y(t)$) to real points of ${\mathcal C}$.
    However, in the sequel we will provide an alternative way for carrying out this computation,
    that allows to certify the existence or non-existence of this kind of points. For this purpose, we denote a
complex value of the parameter $t=u+iv$, where $i^2=-1$ and
$u,v\in {\Bbb R}$, and we represent the complex modulus as $|
\cdot |$. Also, we write
    \[
    \displaystyle{\frac{p_1(u+iv)\cdot \overline{q_1(u+iv)}} {| q_1|^2}}=\displaystyle{\frac{1}{| q_1|^2}}\cdot \left(a(u,v)+ib(u,v)\right)\]
and
\[
\displaystyle{\frac{p_2(u+iv)\cdot \overline{q_2(u+iv)}}{| q_2|^2}}=\displaystyle{\frac{1}{| q_2|^2}}\cdot \left(c(u,v)+id(u,v)\right)
\]
Then the following result, that can be easily verified, holds. Here, we denote
the result of substituting $t=u+iv$ in $\tilde{q}(t)=\mbox{lcm}(q_1,q_2)$, as $\tilde{q}(u,v)$.

\begin{lemma} \label{comp-isol}
Let $P_0\in {\mathcal C}\cap {\Bbb R}^2$. Then, $P_0$ is generated
by a complex value of the parameter $t_0=u_0+iv_0$ if and only if
there exists $w_0\in {\Bbb R}$ satisfying that $(u_0,v_0,w_0)$ is
a real solution of the system
\begin{equation}\label{sistema_pa_2d}
\left\{\begin{array}{c}
b(u,v)=0\\
d(u,v)=0\\
v \cdot |\tilde{q}(u,v)| ^2\cdot w-1=0\end{array}\right.
\end{equation}
\end{lemma}


In order to certify the number of real solutions of System
(\ref{sistema_pa_2d}) we apply Hermite's method (see for example
\cite{cox}). However, these solutions include the complex values
of the parameter generating real points that are also reached by
real values of the parameter. In order to identify the existence
of those solutions, we compute, also by Hermite's method, the
number of real solutions of the system obtained by adding the
following equations to System (\ref{sistema_pa_2d}):
\begin{equation}\label{sistema_pb_2d}
\left\{\begin{array}{c}
x(t)=\displaystyle{\frac{a(u,v)}{| q_1(u,v)|^2}}\\
y(t)=\displaystyle{\frac{b(u,v)}{| q_2(u,v)|^2}}\\
v \cdot \tilde{q}(t)  \cdot w-1=0\end{array}\right.
\end{equation}

So, real isolated points of ${\mathcal C}$ correspond to solutions of System
(\ref{sistema_pa_2d}) which are not solutions of System
(\ref{sistema_pb_2d}).



\subsection{Edges of the Graph.} \label{edges}

In this section, we address the problem of connecting the vertices
of ${\mathcal G}$ (to compute the edges of the graph). For this
purpose, the idea is to introduce between two consecutive critical
lines an intermediate ``non-critical" line, and to connect the
points of ${\mathcal C}$ on each ``non-critical" line with the
points of ${\mathcal C}$ on the critical line immediately on its
right or on its left. In order to do this, we take advantage of
the fact that a parametrization of the curve is available, and we
connect the points just by comparing the parameters generating the
points in the two vertical lines (one of them critical, and the
other one ``non-critical"). The idea is made precise in the
following theorem. Here, we will consider $P_{\infty}$ as
``generated" by both $+\infty$ and $-\infty$, besides other real
values that may also generate it; as usual, $-\infty$ (resp.
$+\infty$) is considered less (resp. greater) than any other real
number compared with it, and $-\infty<+\infty$. This result is
illustrated by Figure 1.

\begin{theorem} \label{th-connect}
Let $x_a,x_b\in {\Bbb R}$ satisfying that: (i) $x_a< x_b$ (resp.
$x_a>x_b$); (ii) there is no critical line $x=x_c$ such that
$x_a\leq x_c<x_b$ (resp. $x_a\geq x_c>x_b$). Also, let $P_a$ be a
real point of ${\mathcal C}$ lying on the line $x=x_a$, generated
by $t_a\in {\Bbb R}$, and let ${\mathcal
V}_b=\{t_{b,1},\ldots,t_{b,n_b},\}$ (including $-\infty,+\infty$,
if $P_{\infty}$ belongs to the line $x=x_b$) be the set of real
values generating the real points of ${\mathcal C}\cap \{x=x_b\}$.
The following statements are true:
\begin{itemize}
\item [(1)] If $x'(t_a)>0$, then $P_a$ must be connected with the point $P_b$ of ${\mathcal C}\cap \{x=x_b\}$ generated by
the least (resp. greatest) element of ${\mathcal V}_b$ which is
greater (resp. less) than $t_a$.
    \item [(2)] If $x'(t_a)<0$, then $P_a$ must be connected with the point $P_b$ of ${\mathcal C}\cap \{x=x_b\}$ generated by the greatest (resp. least) element of ${\mathcal V}_b$ which is less (resp. greater) than $t_a$.
\end{itemize}
\end{theorem}

{\bf Proof.} We prove (1) for the case when $x_a< x_b$; the proofs
of (1) for the case $x_a>x_b$, and of (2) in both cases, are
analogous. Now let $t_c\in {\mathcal V}_b$ be the least element of
${\mathcal V}_b$ which is greater than $t_a$. Since by hypothesis
${\mathcal C}$ has no vertical asymptotes, then $P_a$ must be
connected either with exactly one real point of ${\mathcal C}\cap
\{x=x_b\}$ generated by a real value of the parameter, or with
$P_{\infty}$. Now, we distinguish two different cases, depending
on whether $P_{\infty}$ belongs to the line $x=x_b$, or not. We
begin with the case when $P_{\infty}$ does not belong to $x=x_b$.
So, $P_a$ is connected with a point of ${\mathcal C}\cap
\{x=x_b\}$ generated by some $\tilde{t}\in {\mathcal V}_b$. First of
all, observe that $\tilde{t}>t_a$. Indeed, by hypothesis
${\mathcal C}$ has no vertical asymptotes. Then, $x(t)$ is defined
for every $t$ between $t_a$ and $\tilde{t}$, and since $x(t)$
is a quotient of polynomials, $x'(t)$ is also differentiable
there. Moreover, $x'(t)$ cannot vanish between $t_a$ and
$\tilde{t}$ because by hypothesis there does not exist any
critical line between $x=x_a$ and $x=x_b$. Hence, the sign of
$x'(t)$ is constant in the interval lying between $t_a$ and
$\tilde{t}$, and since $x'(t_a)>0$, then $x'(t)>0$ in that interval; therefore,
$x(t)$ is increasing there. So, since $x_a=x(t_a)<x(\tilde{t})$
we deduce that $t_a<\tilde{t}$.

Now our aim is to prove that $\tilde{t}=t_c$. For this purpose,
 observe that $\tilde{t}\geq t_c$ because $t_c$ is
the least element of ${\mathcal V}_b$ greater than $t_a$; hence,
we just have to prove that $\tilde{t}>t_c$ cannot occur. Assume by
contradiction that $\tilde{t}>t_c$. Since
$x(t_c)=x_b=x(\tilde{t})$ and $x(t)$ is differentiable along
$[t_a,\tilde{t})$, by Rolle's Theorem $x'(t)$ must vanish at some
point of $(t_c,\tilde{t})$. However, this is absurd because
$x'(t)$ is strictly positive in $[t_a,\tilde{t})$, which contains
$(t_c,\tilde{t})$.

Finally, let
us consider the case when $P_{\infty}$ belongs to the line
$x=x_b$. If there exists $\hat{t}\in {\mathcal V}_b$, $\hat{t}\neq
+\infty$, with $\hat{t}>t_a$, then $P$ must be connected with
$\hat{P}=\varphi(\hat{t})$, since otherwise by adapting the above
argument one reaches a
contradiction. On the other hand,
if $t_a$ is greater than every real element of ${\mathcal V}_b$,
then $P_a$ cannot be connected with any other point of ${\mathcal
C}\cap \{x=x_b\}$ but $P_{\infty}$; however, since we consider
$P_{\infty}$ generated by $+\infty$, and $t_a<\infty$, the rule also
holds in this case. \qed


\begin{figure}[ht]
\begin{center}
\centerline{$\begin{array}{ccc}
\psfig{figure=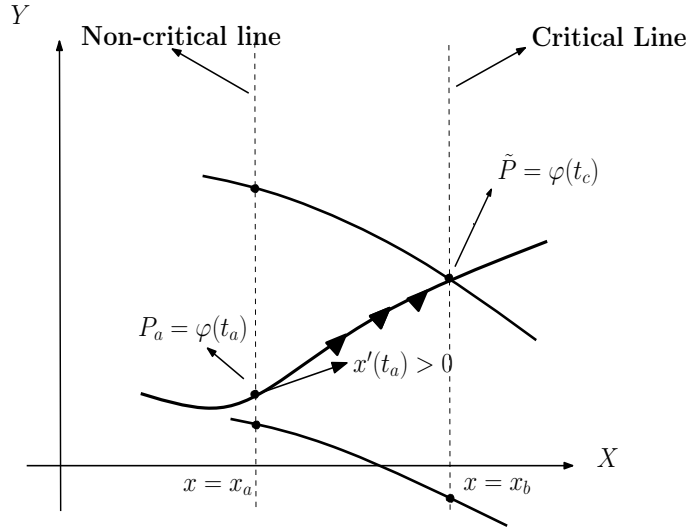,width=9cm,height=7cm}
\end{array}$}
\end{center}
\caption{Connecting Points}
\end{figure}

\subsection{Full Algorithm} \label{algorit-planar}

The following algorithm {\tt Planar-Top} can be derived from the preceding subsections.

\underline{\tt Planar-Top Algorithm:}

{\sf Input:}  a planar curve ${\mathcal C}$, parametrized by
\[\varphi(t)=(x(t),y(t))=\left(\displaystyle{\frac{p_1(t)}{q_2(t)}},\displaystyle{\frac{p_2(t)}{q_2(t)}}\right),\]
fulfilling: (i) $p_i(t),q_i(t)\in {\Bbb Z}[t]$ for $i=1,2$,
$\gcd(p_i,q_i)=1$ for $i=1,2$; (ii) $\varphi(t)$ is proper; (iii)
${\mathcal C}$ has no vertical asymptotes.

{\sf Output:} a planar graph ${\mathcal G}$ homeomorphic to the
curve.

\begin{itemize}
\item [(1)] (Critical Points) Compute the polynomial $m(t)$ in Corollary \ref{gen-critical}, and the real roots of $m(t)$. Then, compute:
 \begin{itemize}
 \item [(1.1)] The critical points of ${\mathcal C}$ (by evaluating $x(t),y(t)$ at the real roots of $m(t)$). Store these points in a list \[{\mathcal L}_{crit}=[P_1, \ldots,P_r],\]For each of these points, store its coordinates, and the list of real $t$-values generating them.
 \item [(1.2)] The point $P_{\infty}=(x_{\infty},y_{\infty})$ (if it exists), and the list of $t$-values generating it.
 \end{itemize}
\item [(2)] (Points of ${\mathcal C}$ on Critical Lines)
\begin{itemize}
\item [(2.1)] For $\ell$ from 1 to $r$, compute the real points of ${\mathcal C}$ lying on the line $x=x_{\ell}$. Store these points in a list
    \[{\mathcal L}_{\ell}=[P_{\ell,1},\ldots,P_{\ell,j_{\ell}}]\]For each of these points, store its coordinates, and the list of real $t$-values generating them.
\item [(2.2)] Check whether $P_{\infty}$ belongs to some of the above lines $x=x_{\ell}$. In the affirmative case, go to (3); otherwise,
compute the real points of ${\mathcal C}$ lying on $x=x_{\infty}$. Store these points in a list
\[{\mathcal L}_{\infty}=[P_{\infty,1},\ldots,P_{\infty,m}]\]For each of these points, store its coordinates, and the list of real $t$-values generating them.
    \end{itemize}
    \item [(3)] (Points of ${\mathcal C}$ on Non-Critical lines)
    \begin{itemize}
    \item [(3.1)] Let ${\mathcal N}=\{a_1\ldots,a_{s}\}$, $a_1<\ldots<a_{s}$, be the set consisting
    of the $x$-coordinates of the critical points computed in (1.1) and (1.2). Also, let $\bar{a}_0=a_1-1$, $\bar{a}_{s}=a_s+1$,
    and for $i$ from 1 to
    $s-1$ let $\bar{a}_i=(a_i+a_{i+1})/2$.
        \item [(3.2)] For $j$ from 0 to $s$, compute the real points of ${\mathcal C}$ lying on the line $x=\bar{a}_j$; store these points in a list
            \[{\mathcal N}_j=[\bar{P}_{j,1},\ldots,\bar{P}_{j,\alpha_j}]\]For each of these points, store its coordinates, and the list of real $t$-values generating them.
        \end{itemize}
    \item [(4)] (Edges)
    \begin{itemize}
    \item [(4.1)] For $i$ from 0 to $s-1$, connect the points of ${\mathcal C}$ lying on $x=\bar{a}_i$ and $x=a_{i+1}$ by applying Theorem \ref{th-connect}.
        \item [(4.2)] For $i$ from 1 to $s$, connect the points of ${\mathcal C}$ lying on $x=\bar{a}_i$ and $x=a_i$ by applying Theorem \ref{th-connect}.
            \end{itemize}
            \item [(5)] (Isolated vertices) Compute the real isolated points of the curve, and add them to the graph.
\end{itemize}

We will provide several examples of the output of the algorithm in Section \ref{Examp}.

\section{Computation of the Graph Associated with a Space Curve}\label{sec-space}

In this section we let ${\mathcal C}\subset {\Bbb R}^3$ be a real curve, parametrized by
\[\varphi(t)=(x(t),y(t),z(t))=\left(\displaystyle{\frac{p_1(t)}{q_1(t)}},\displaystyle{\frac{p_2(t)}{q_2(t)}},\displaystyle{\frac{p_3(t)}{q_3(t)}}\right)\]
where $\gcd(p_1,q_1)=\gcd(p_2,q_2)=\gcd(p_3,q_3)=1$. In the sequel, we consider the problem of algorithmically computing a graph ${\mathcal G}$ homeomorphic to ${\mathcal C}$. In order to do that, we will follow the strategy used to address the implicit case in \cite{JG-Sendra}, \cite{Diat}, \cite{ElKa}; more precisely, we need to perform the following steps:
\begin{itemize}
\item [(1)] Project the curve onto a coordinate plane (the $xy$-plane, in our case)
 \item [(2)] Compute the graph $\overline{\mathcal G}$ associated with the projection (by using the algorithm given in Section \ref{plane-case})
     \item [(3)] Lift the graph $\overline{\mathcal G}$ of the projection, to get ${\mathcal G}$.
     \end{itemize}
     As in \cite{Diat} and \cite{ElKa}, here we will require just one projection
     in order to perform the lifting phase. Now in the following subsections we first describe the hypotheses that we request on the input curve (essentially, that it is properly parametrized, and that it is ``correctly placed" in space); then, we present the ideas and results needed for computing the vertices and edges of the graph, and finally we provide the full algorithm.

\subsection{Hypotheses}\label{hyp-3d}

Since $\mathcal C$ is rational, if it is not a line parallel to
the $z$-axis, then its projection onto the $xy$ plane, denoted as
$\pi_{xy}({\mathcal C})$, is an algebraic rational curve and can
be parametrized by
\[\psi(t)=(x(t),y(t))=\left(\displaystyle{\frac{p_1(t)}{q_1(t)}},\displaystyle{\frac{p_2(t)}{q_2(t)}}\right)\]
Thus, in the sequel we assume that the following hypotheses hold:

\begin{itemize}
\item [(i)] ${\mathcal C}$ has no asymptotes parallel to the $z$-axis (in particular, it is not normal to the $xy$-plane).
\item [(ii)] $\psi(t)$ is a proper parametrization of $\pi_{xy}({\mathcal C})$.
\item [(iii)] $\pi_{xy}({\mathcal C})$ has no asymptotes parallel to the $y$-axis.
\end{itemize}

In particular, hypotheses (ii) and (iii) imply that the graph of
$\pi_{xy}({\mathcal C})$ can be computed by using the Planar-Top
Algorithm. Now if the  parametrization $\varphi(t)$ of ${\mathcal
C}$ is not proper, then $\psi(t)$ cannot be a proper
parametrization of $\pi_{xy}({\mathcal C})$ either; then, in
particular (ii) implies that ${\mathcal C}$ is properly
parametrized. However, the converse does not necessarily hold,
i.e. it can happen that $\varphi(t)$ is proper, but $\psi(t)$ is
not. From Section \ref{plane-case}, we know how to check
hypotheses (ii) and (iii). In order to check hypothesis (i), the
next lemma, analogous to Lemma \ref{asymp-plane}, can be applied.

\begin{lemma} \label{asint-space}
${\mathcal C}$ has an asymptote parallel to the $z$-axis iff one of the following conditions happen: (a) $q_3(t)$ has some real root, which is not a real root of $q_1(t)\cdot q_2(t)$; (b) $\deg(p_3))>\deg(q_3)$ but $\deg(p_2)\leq \deg(q_2)$, $\deg(p_1)\leq \deg(q_1)$.
\end{lemma}

Moreover, hypothesis (i) implies the following relationship between the points $Q_{\infty}=\lim_{t\to \pm \infty}\psi(t)$, $P_{\infty}=\lim_{t\to \pm \infty}\varphi(t)$. Recall from Section \ref{sec-background} that they are the only points of $\pi_{xy}({\mathcal C})$ and ${\mathcal C}$, respectively, that {\it may} not be reached by any complex value of the parameter.

\begin{lemma} \label{inf-points}
Assume that hypothesis (i) holds. Then, $P_{\infty}$ exists iff $Q_{\infty}$ exists, and $\pi_{xy}(P_{\infty})=Q_{\infty}$.
\end{lemma}

{\bf Proof.} If $P_{\infty}$ exists, then it is clear that
$Q_{\infty}$ exists and is the projection of $P_{\infty}$.
Conversely, if $Q_{\infty}=(x_{\infty},y_{\infty})\in {\Bbb R}^2$
then $P_{\infty}$ exists because  ${\mathcal C}$ has no
asymptotes. \qed

On the other hand, hypothesis (ii) leads to the following result. Here, the notion of {\it birationality} arises; essentially, the projection of ${\mathcal C}$ is said to be {\it birational} if there are not two different branches of ${\mathcal C}$ whose projections overlap (see Chapter 5 in \cite{cox-1} for further information on birationality).



\begin{theorem} \label{lem-third-hyp}
Assume that ${\mathcal C}$ is not a line parallel to the $z$-axis. Then, if $\psi(t)$ is proper, the projection of ${\mathcal C}$ onto the $xy$-plane is birational. Conversely, if $\varphi(t)$ is proper and the projection of ${\mathcal C}$ onto the $xy$-plane is birational, then $\psi(t)$ is proper.
\end{theorem}

{\bf Proof.} Let us see $(\Rightarrow)$. For this purpose, let $Q\in \pi_{xy}({\mathcal C})$, $Q\neq Q_{\infty}$, satisfying that there are at least two different points $P,\tilde{P}\in {\mathcal C}$ projecting onto $Q$. Since $Q\neq Q_{\infty}$, by Lemma \ref{inf-points} none of these points is $P_{\infty}$, and hence both are reached by $\varphi(t)$. Let $t_p\neq \tilde{t}_p$ be the $t$-values generating $P,\tilde{P}$, respectively. Then, $\psi(t_p)=\psi(\tilde{t}_p)$, and thus $Q$ is generated by two different values of the parameter. But since $\psi(t)$ is proper, this can only happen for finitely many points, and thus the projection is birational. Conversely, given any $Q\in \pi_{xy}({\mathcal C})$, $Q\neq Q_{\infty}$, not generated by any root of $q_3(t)$ (notice that we are excluding finitely many points), the $t$-values reaching $Q$ are exactly those ones generating the points of ${\mathcal C}$ that are projected onto $Q$. Since $\varphi(t)$ is proper, almost all points of ${\mathcal C}$ are generated by just one value of the parameter. And since the projection is birational, we conclude that almost all points of $\pi_{xy}({\mathcal C})$ come from just one point of ${\mathcal C}$, and therefore almost all points of $\pi_{xy}({\mathcal C})$ are generated by just one value of the parameter. So, $(\Leftarrow)$ holds. \qed

It is well-known that almost all affine transformations of the
type $\{X=x+az,y=y+bz,z\}$ transform ${\mathcal C}$
so that its $xy$-projection is birational. So,
if $\varphi(t)$ is proper, almost all of these transformations set
${\mathcal C}$ proper. Moreover, if $\varphi(t)$ is not proper
there exist reparametrization algorithms (see \cite{Tomas},
\cite{Seder}). Therefore, in the sequel we will assume that the
above hypotheses hold.

\subsection{Definition of the Space Graph.}\label{graph-3d}

By assuming the hypotheses of the preceding subsection hold, we
can compute the graph $\overline{\mathcal G}$ associated with
$\pi_{xy}({\mathcal C})$ with the Planar-Top Algorithm described
in Section \ref{plane-case}. Hence, in the following we will
assume that this process has already been carried out.

Now we make precise the definition of the graph ${\mathcal G}$
that we want to compute.

\begin{definition} \label{graph-assoc-3d}
Let ${\mathcal C}$ be a space curve in the above conditions. Then, the {\sf graph associated with ${\mathcal C}$}, ${\mathcal G}$, is the following graph:
\begin{itemize}
\item [(i)] Its vertices are the real points of ${\mathcal C}$ giving rise (by projection) to the vertices of $\overline{\mathcal G}$.
    \item [(ii)] Its edges are the result of ``lifting" to space the edges of $\overline{\mathcal G}$, i.e. of computing, for each edge $\ell$ of $\overline{\mathcal G}$, an space segment $\ell'$ corresponding to the branch of ${\mathcal C}$ giving rise (by projection) to $\ell$.
        \end{itemize}
        \end{definition}

        Hence, we have to lift to space the vertices and edges of $\overline{\mathcal G}$ in order to compute ${\mathcal G}$. Let us see that this lifting operation is well-defined.

\begin{theorem} \label{th-vert-lift}
Every vertex of $\overline{\mathcal G}$, except perhaps the isolated vertices, lifts to at least one real space point of ${\mathcal C}$.
\end{theorem}

{\bf Proof.} Every point of $Q\in \pi_{xy}({\mathcal C})\cap {\Bbb
R}^2$ fulfills one of the following conditions: (1)
$Q=Q_{\infty}$; (2) there exists $t_0\in {\Bbb R}$ satisfying that
$Q=\psi(t_0)$; (3) $Q$ does not fulfill (2), but there exists
$t_0\in {\Bbb C}$ such that $Q=\psi(t_0)$. In the first case,$Q$
is lifted to $P_{\infty}$ by Lemma \ref{inf-points}.  If $Q$
belongs to the second group, then it is lifted to $P=\varphi(t_0)$
because ${\mathcal C}$ has no asymptotes parallel to the $z$-axis.
Finally, if $Q$ belongs to the third group then it is an isolated
point of $\pi_{xy}({\mathcal C})$; in this case, $Q$ comes from a
real point of ${\mathcal C}$ iff $z(t_0)\in {\Bbb R}$. \qed

\begin{remark} \label{rem-isol}
Real isolated points of $\pi_{xy}({\mathcal C})$ may come from real isolated points of ${\mathcal C}$, or from points of ${\mathcal C}$ whose $z$-coordinate is complex. In any case, thanks to hypothesis (i) they do not come from branches of ${\mathcal C}$ normal to the $xy$-plane.
\end{remark}

Now let us consider the lifting of the edges of the planar graph.
The next result guarantees that, under the considered hypotheses,
this lifting process can be always carried out. In particular, it
implies that there are no real branches of $\pi_{xy}({\mathcal
C})$ coming from complex components of ${\mathcal C}$ (which is a
phenomenon that in general can happen when working with space
algebraic curves; see for example p. 734 in \cite{JG-Sendra}).

\begin{theorem} \label{th-edges-lift}
Under the considered hypothesis, for every edge $\ell$ of $\overline{\mathcal G}$ there exists one and just one branch of ${\mathcal C}$ giving rise to $\ell$.
\end{theorem}

{\bf Proof.} Let $\ell$ be an edge of $\overline{\mathcal G}$. By
construction of the graph provided in Section \ref{plane-case}, ,
if $Q_{\infty}$ exists, it is always included as a vertex of
$\overline{\mathcal G}$. So there exists a real open interval
$I\subset {\Bbb R}$ such that $\psi(I)$ generates the real branch
of $\pi_{xy}({\mathcal C})$, that we denote by ${\mathcal L}$,
corresponding to $\ell$. On the other hand, for every $t\in I$ we
have that $z(t)$ must be defined, because otherwise ${\mathcal C}$
has an asymptote parallel to the $z$-axis. Then, $\varphi(t)$ is
defined for every $t\in I$, and gives rise to a real connected
branch of ${\mathcal C}$ projecting as ${\mathcal L}$.
Furthermore, since $\varphi(t)$ is proper by hypothesis, the
projection onto the $xy$-plane is birational by Theorem
\ref{lem-third-hyp}. Hence, there are just finitely many points of
${\mathcal C}$ giving rise, by projection, to the same point of
$\pi_{xy}({\mathcal C})$; but none of these points can give rise
to a point of ${\mathcal L}$, because such a point would create a
singularity of $\pi_{xy}({\mathcal C})$ which would split $\ell$
into two different edges, and $\ell$ is already an edge of
$\overline{\mathcal G}$. Then, we conclude that ${\mathcal L}$
lifts to a unique connected real branch of ${\mathcal C}$. \qed

\subsection{Computation of the Vertices}\label{vertices-3d}
From Definition \ref{graph-assoc-3d}, this process is the lifting of the vertices of
$\overline{\mathcal G}$. From the construction of the planar
graph, one may see that for each vertex $Q_i=(x_i,y_i)$ of
$\overline{\mathcal G}$ the algorithm stores the real values
$t_{i,1},\ldots,t_{i,r}$ of the parameter generating it. For a
fixed $i$, $z(t_{i,j})$ is well-defined for $j\in \{1,\ldots,r\}$,
since otherwise ${\mathcal C}$ has an asymptote parallel to the
$z$-axis. Hence, $Q_i$ is lifted to the space points
\[P_{i,1}=\varphi(t_{i,1}),\ldots,P_{i,r}=\varphi(t_{i,r})\] 
Furthermore, if $Q_{\infty}$ exists, then it is lifted to $P_{\infty}$ and to the space points reached by
the real values of the parameter generating $Q_{\infty}$, if any. Proceeding this way, the only remaining
space vertices are the real isolated ones (which, by Proposition \ref{R-normal}, are generated
by complex values of the parameter). So, in the rest of the subsection we consider this kind of points.

From Theorem \ref{th-vert-lift}, the real isolated vertices of
$\pi_{xy}({\mathcal C})$ do not necessarily come from real
isolated points of ${\mathcal C}$ (since they may be the
projection of complex space points). Conversely, a real isolated
point of ${\mathcal C}$ does not necessarily project as an
isolated point of $\pi_{xy}({\mathcal C})$, because its projection
may coincide with the projection of some other real point of
${\mathcal C}$ which is not isolated. However, the next result
ensures that isolated points of ${\mathcal C}$ always project as
vertices of $\overline{\mathcal G}$; therefore, these points are
computed when lifting the planar vertices.

\begin{lemma} \label{lemma-isol}
Let $P\in {\mathcal C}$ be a real isolated point. Then, $\pi_{xy}(P)$ is a vertex of $\overline{\mathcal G}$.
\end{lemma}

{\bf Proof.} If $\pi_{xy}(P)$ is an isolated point of
$\pi_{xy}({\mathcal C})$, then the statement is true. Otherwise,
there exists a point $P'\neq P$ in a real branch of ${\mathcal C}$
such that $\pi_{xy}(P)=\pi_{xy}(P')$. Observe that $P$ cannot be
$P_{\infty}$ because it is isolated. Therefore, suppose that  it
is reached via $\varphi(t)$ by $t_p\in {\Bbb C}$. Now we
distinguish the cases $P'\neq P_{\infty}$ or $P'= P_{\infty}$,
respectively. If $P'\neq P_{\infty}$, then $P'=\varphi(t_{p'})$
with $t_{p'}\in {\Bbb R}$. Thus, $\pi_{xy}(P)$ is generated via
$\varphi(t)$ by two different values of the parameter, namely
$t_p,t_{p'}$, and since $\varphi(t)$ is proper, $\pi_{xy}(P)$ is a
self-intersection of $\pi_{xy}({\mathcal C})$. Hence, it is a
singularity of $\pi_{xy}({\mathcal C})$, and the statement
follows. Finally, if $P'= P_{\infty}$ then
$\pi_{xy}(P')=Q_{\infty}$ and therefore it is also a vertex of
$\overline{\mathcal G}$. \qed

Then, we might recover isolated singularities of ${\mathcal C}$ by determining the complex values of the parameter that generate (by projection) vertices of $\overline{\mathcal G}$, and by computing those real points of ${\mathcal C}$ which are generated by those values. Nevertheless, in the sequel we consider an alternative method, analogous to that in Subsection \ref{vertices}. For this purpose, the following lemma is needed. Here, we denote a complex value of the parameter $t$ as $t=u+iv$, where $i^2=-1$ and $u,v\in {\Bbb R}$. Also, we write
\[
\begin{array}{c}
\displaystyle{\frac{p_1(u+iv)\cdot \overline{q_1(u+iv)}}{| q_1|^2}}=\displaystyle{\frac{1}{| q_1|^2}}\cdot \left(a(u,v)+ib(u,v)\right)\\
\displaystyle{\frac{p_2(u+iv)\cdot \overline{q_2(u+iv)}}{| q_2|^2}}=\displaystyle{\frac{1}{| q_2|^2}}\cdot \left(c(u,v)+id(u,v)\right),
\end{array}
\]
and
\[
\displaystyle{\frac{p_3(u+iv)\cdot \overline{q_3(u+iv)}}{| q_3|^2}}=\displaystyle{\frac{1}{| q_3|^2}}\cdot \left(e(u,v)+ih(u,v)\right)
\]
Then the following
result, analogous to Lemma \ref{comp-isol}, holds. Here, $\tilde{q}(u,v)$ denotes the result of substituting $t=u+iv$ in $\mbox{lcm}(q_1,q_2,q_3)$. As in Lemma \ref{comp-isol}, by applying the following result one computes a finite set of complex points which contains the complex points generating the isolated singularities of the space curve.

\begin{lemma} \label{isol-points-3d}
{Let $P\in {\mathcal C}\cap {\Bbb R}^3$. Then, $P$ is generated by
a complex value of the parameter $t_0=u_0+iv_0$ if and only if
there exists $w_0\in {\Bbb R}$ satisfying that $(u_0,v_0,w_0)$ is
a real solution of the system }

\begin{equation}\label{sistema}
\left\{ \begin{array}{c}
b(u,v)=0\\
d(u,v)=0\\
h(u,v)=0\\
v \cdot |\tilde{q}(u,v)|^2 \cdot w-1=0
\end{array} \right.
\end{equation}
\end{lemma}

As in the planar case, one can certify the number of real
solutions of the system by Hermite's method; also, one can
construct another system whose solutions correspond to complex
values of the parameter generating points that are also reached by
real values of the parameter, and proceed as in
the 2D case.

\subsection{Computation of the Edges}\label{edges-3d}

The method consists of the lifting of the edges of
$\overline{\mathcal G}$. So, let $\ell$ be an edge of
$\overline{\mathcal G}$; by Theorem \ref{th-edges-lift}, $\ell$ is
lifted to an space edge $\ell' \in {\mathcal G}$. In order to
compute $\ell'$, the crucial observation is that the computation
of the edges of $\overline{\mathcal G}$ is in fact done by
connecting not points, but values of the parameter $t$. Hence,
each edge $\ell$ can be identified with a pair \[ [t_a,\tilde{t}]
\]
where $t_a,\tilde{t}$ belong to ${\Bbb R}\cup
\{-\infty,+\infty\}$, and where the vertices of
$\overline{\mathcal G}$ defining $\ell$ are $Q_a=\psi(t_a)$,
$\tilde{Q}=\psi(\tilde{t})$ (see also Figure 1); here,
$Q_{\infty}=\psi(\pm \infty)$. Hence, $\ell$ is lifted to the
space segment connecting the points $P_a=\varphi(t_a)$,
$\tilde{P}=\varphi(\tilde{t})$; also, $P_{\infty}=\varphi(\pm
\infty)$. Notice that this idea works perfectly when $\tilde{Q}$
is the projection of several real points of ${\mathcal C}$.

\subsection{Full Algorithm}

The following algorithm {\tt Space-Top} can be derived from the preceding subsections.

\underline{\tt Space-Top Algorithm:}

{\sf Input:} a space curve ${\mathcal C}$, parametrized by \[\varphi(t)=(x(t),y(t),z(t))=\left(\displaystyle{\frac{p_1(t)}{q_2(t)}},\displaystyle{\frac{p_2(t)}{q_2(t)}},\displaystyle{\frac{p_3(t)}{q_3(t)}}\right),\]
fulfilling: (i) $p_i(t),q_i(t)\in {\Bbb Z}[t]$ for $i=1,2,3$, $\gcd(p_i,q_i)=1$ for $i=1,2,3$; (ii) ${\mathcal C}$ has no asymptotes parallel to the $z$-axis; (iii) $\psi(t)=\left(\displaystyle{\frac{p_1(t)}{q_2(t)}},\displaystyle{\frac{p_2(t)}{q_2(t)}}\right)$ is proper; (iv) $\pi_{xy}({\mathcal C})$ has no asymptotes parallel to the $y$-axis.

{\sf Output:}  a space graph ${\mathcal G}$ homeomorphic to ${\mathcal C}$.

\begin{itemize}
\item [(1)] (Projection) Compute the graph $\overline{\mathcal G}$ associated with the projection $\pi_{xy}({\mathcal C})$ of ${\mathcal C}$ onto the $xy$-plane, parametrized by $\psi(t)$, by applying {\tt Planar-Top}.
    \item [(2)] (Lifting phase)
    \begin{itemize}
    \item [(2.1)] (Vertices) For each vertex of $\overline{\mathcal G}$: if $P$ is generated by $t_1,\ldots,t_p$ where $t_1,\ldots,t_p\in {\Bbb R}\cup \{-\infty,+\infty\}$, then $P$ lifts to the points $\varphi(t_1),\ldots,\varphi(t_p)$;  $Q_{\infty}$, if it exists, lifts to $P_{\infty}$, and we write $P_{\infty}=\varphi(\pm \infty)$.
        \item [(2.2)] (Edges) For each edge of $\overline{\mathcal G}$: if $\ell$ is identified (according to Subsection \ref{edges-3d}) with $[t_a,t_b]$, where $t_a,t_b\in {\Bbb R}\cup \{-\infty,+\infty\}$, then it is lifted to the space edge obtained by connecting $\varphi(t_a),\varphi(t_b)$ by means of a segment.
            \end{itemize}
    \item [(3)] (Isolated vertices) Add to ${\mathcal G}$ the real isolated singularities of ${\mathcal C}$.
    \end{itemize}

 Several examples of the output of this algorithm are presented in the next section.

\section{Experimentation and Examples} \label{Examp}

The algorithm has been implemented in \texttt{Maple 13}, and the examples
run on an Intel Core 2 Duo processor with speeds revving up to
1.83 GHz. The implementation allows the option of computing
isolated points, or not. The reason for introducing this option is
that the number of isolated points is certified by means of
Hermite's method, and this method may be costly.

On the other hand, the user can decide the number of digits used
in the computation. Suppose we denote such a number by $n$. Then,
when running the algorithm, the computing starts using $n$ digits.
However, if the algorithm detects that the number of points in a
vertical line is not the right number, the precision is
automatically increased by 5 more digits and the whole process
starts again. In our experimentations, we usually set $n=10$, the
default value of \texttt{Digits} variable in \texttt{Maple}, and
in the implementation, the number of digits is limited to a
maximum of 500, although we have never needed more than 70 digits.


Next, we first present examples of the 2D algorithm. In Table 1, we include, for each curve, the degree of the
parametrization (i.e. the maximum exponent of the parameter in the
numerators and denominators of the components of the
parametrization, $d_p$), the total degree of the implicit equation
($d_i$), the number of terms of the implicit equation (n.terms),
 the
timings in seconds corresponding to the graph without computing isolated
points ($t_0$) or computing them ($t_1$), and the number of digits
used in the computations. The parametrizations corresponding to
these examples are given in Appendix I.

\begin{center}
\begin{tabular}{|c|l|l|c|c|c|c|c|} \hline
Example & $d_p$ & $d_i$ & n.terms  & $t_0$ & $t_1$ & Digits  \\
\hline \hline 1 & 3 & 6 & 16  & 0.359 & 1.672 & 10  \\
\hline 2 & 8 & 8 & 25  & 0.891 & 1.078 & 10  \\
\hline 3 & 8 & 8 & 9  & 10.250 & 71.172 & 40  \\
\hline 4 & 4 & 4 & 7  & 0.109 & 0.110 & 10 \\
\hline 5 & 6 & 6 & 28  & 0.203 & 11.859 & 10 \\
\hline 6 & 8 & 8 & 21  & 0.171 & 2.50 & 10 \\
\hline 7 & 23 & 23 & 335  & 49.797 & $> 1$ h. & 10 \\
\hline 8 & 6 & 12 & 49  & 13.625 & $> 1$ h. & 10 \\
\hline 9 & 17 & 17 & 171  & 1.656 & $> 1$ h. & 10 \\
\hline
\end{tabular}

{\bf Table 1:} 2D Examples.
\end{center}

One may notice that as the degree increases (see Example 7 or
Example 9) the computation of the isolated points turns very
costly. An alternative for those cases could be to detect isolated
points directly by checking the existence of complex values of the
parameter corresponding to real singular points; users interested in certifying
rigourously the number of isolated singularities, can choose to apply
Hermite's method later.

The pictures corresponding to the examples in Table 1 can be found
in Figure 2; from left to right we have Examples 1, 2, 3 in the
first row, 4, 5, 6 in the second row and 7, 8, 9 in the third one.
Examples 2 and 6 are the offsets of the cardioid and of the
cubical cusp, respectively; furthermore, Example 4 corresponds to
the epitrochoid. Notice that the curves in Examples 2, 3 and 4 are
not
in generic position.

\begin{figure}[ht]
\begin{center}
\centerline{$\begin{array}{ccc}
\psfig{figure=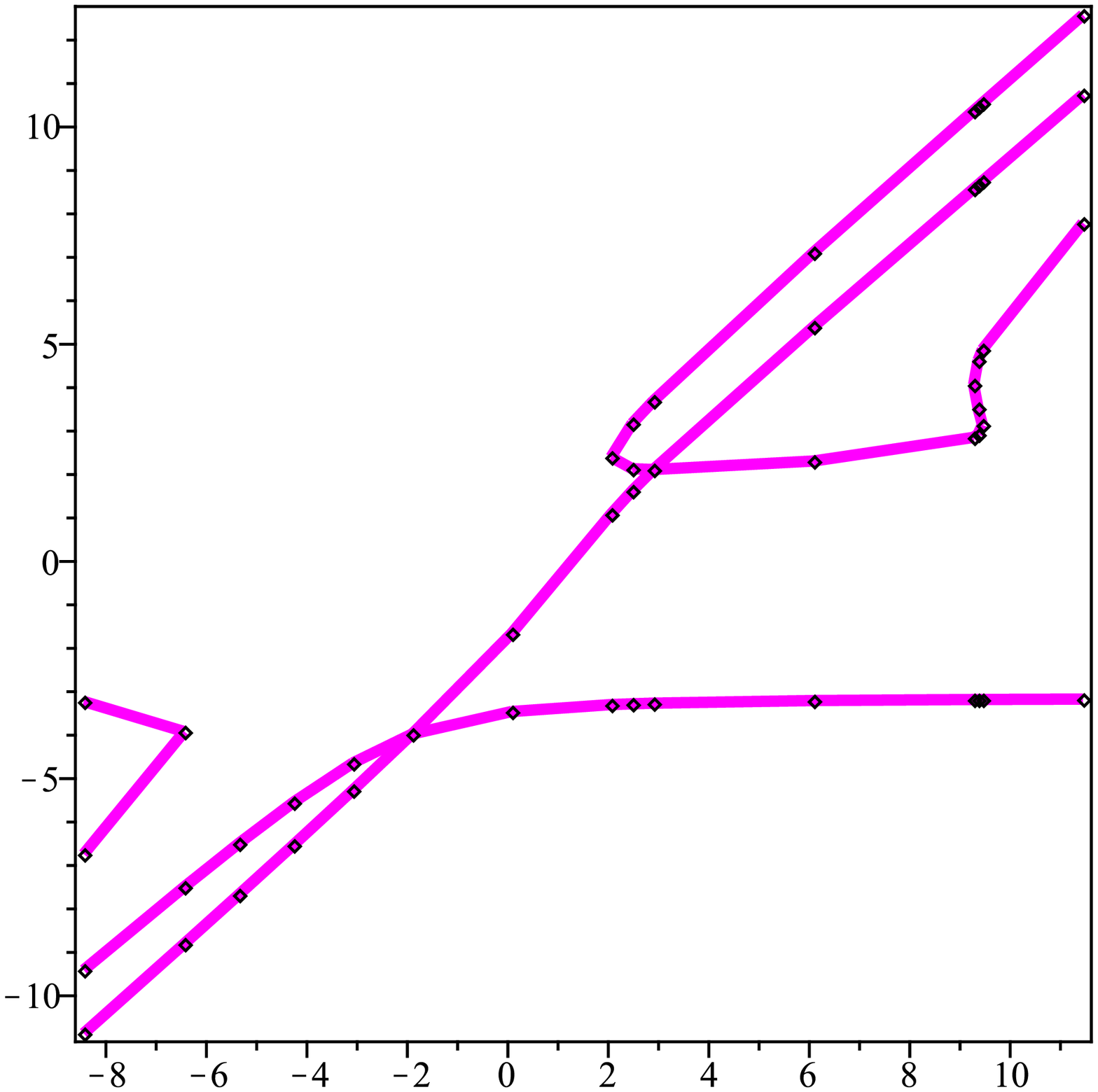,width=5cm,height=5cm} &
\psfig{figure=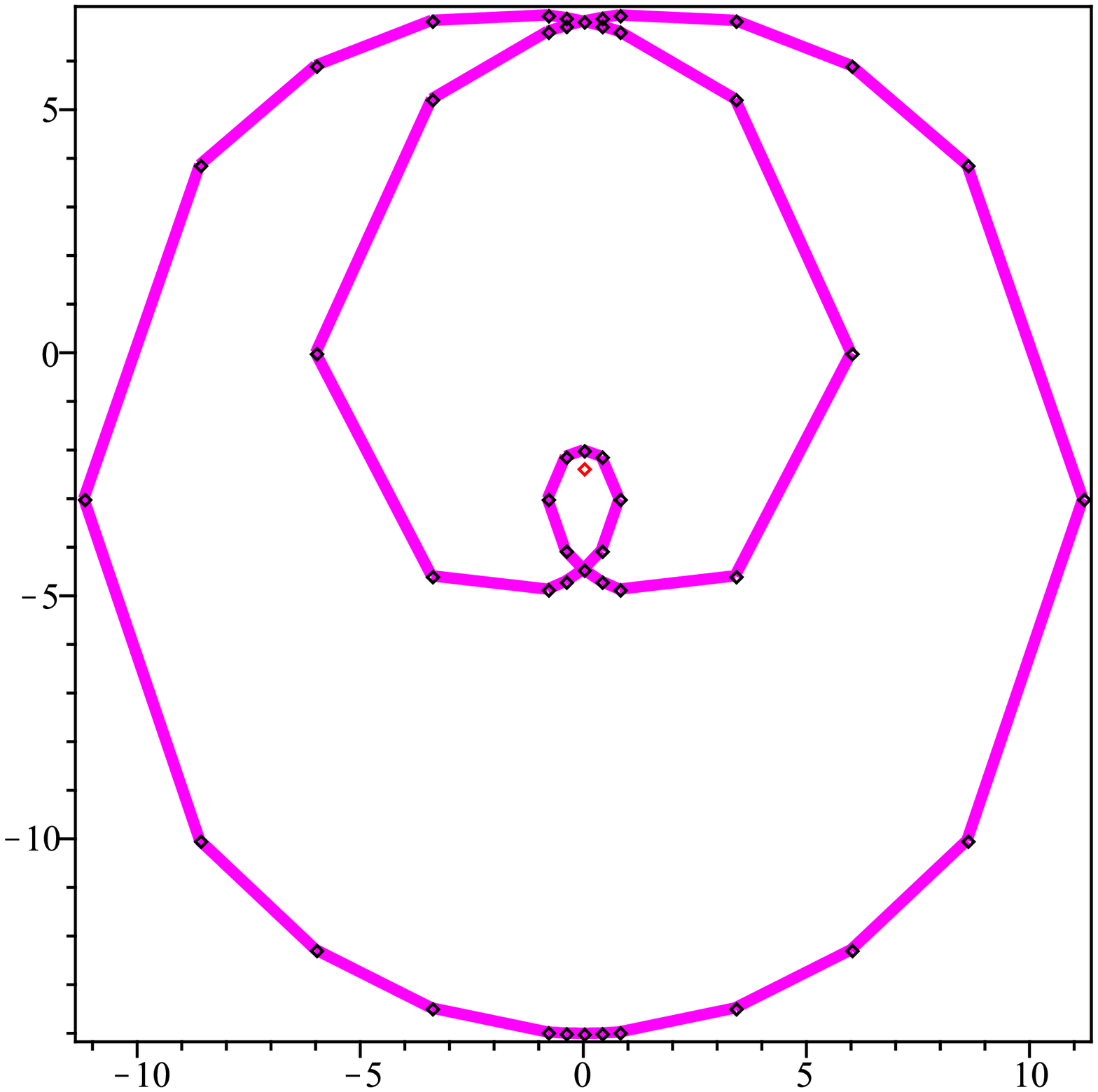,width=5cm,height=5cm} &
\psfig{figure=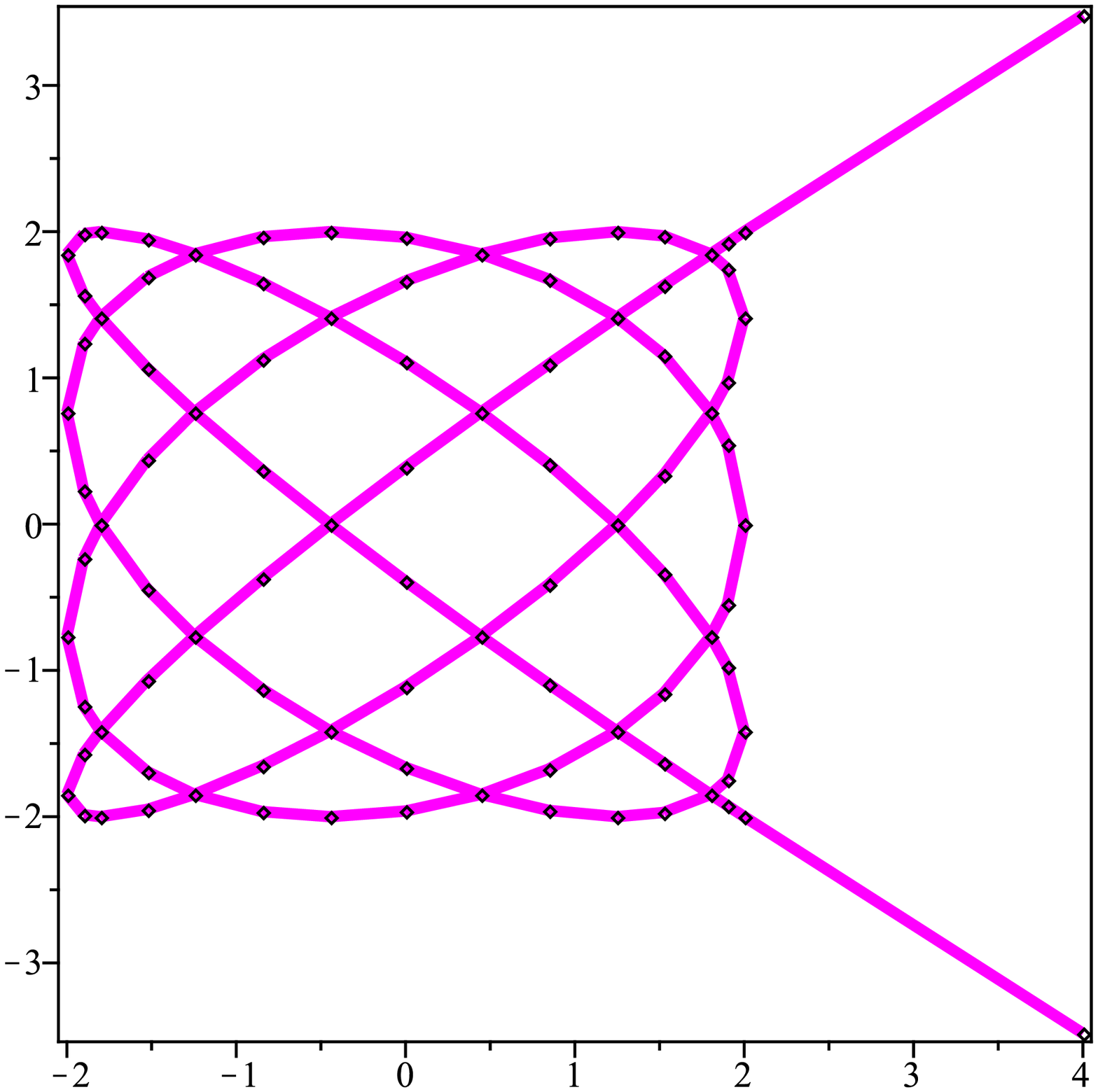,width=5cm,height=5cm} \\
\psfig{figure=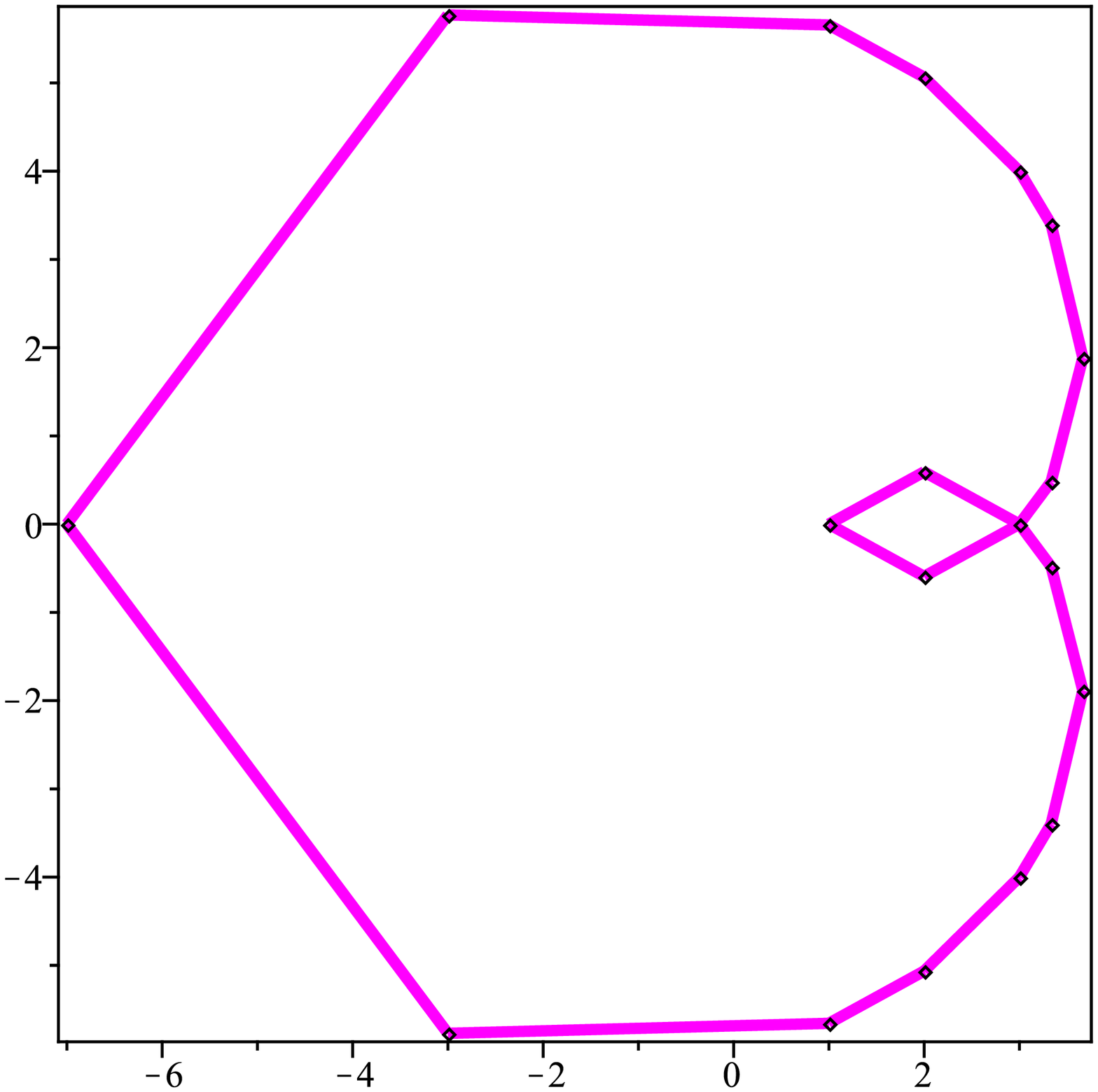,width=5cm,height=5cm} &
\psfig{figure=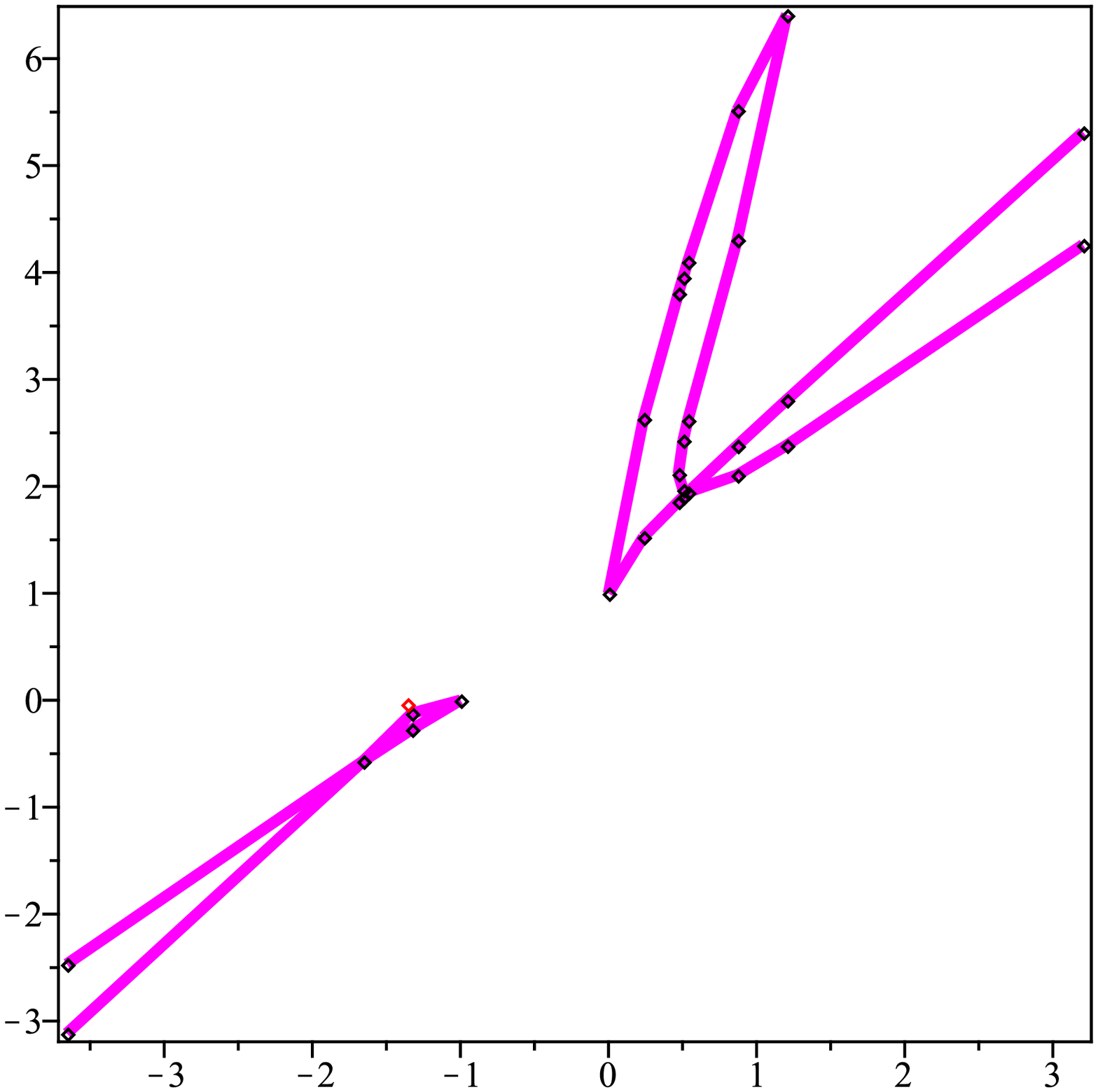,width=5cm,height=5cm} &
\psfig{figure=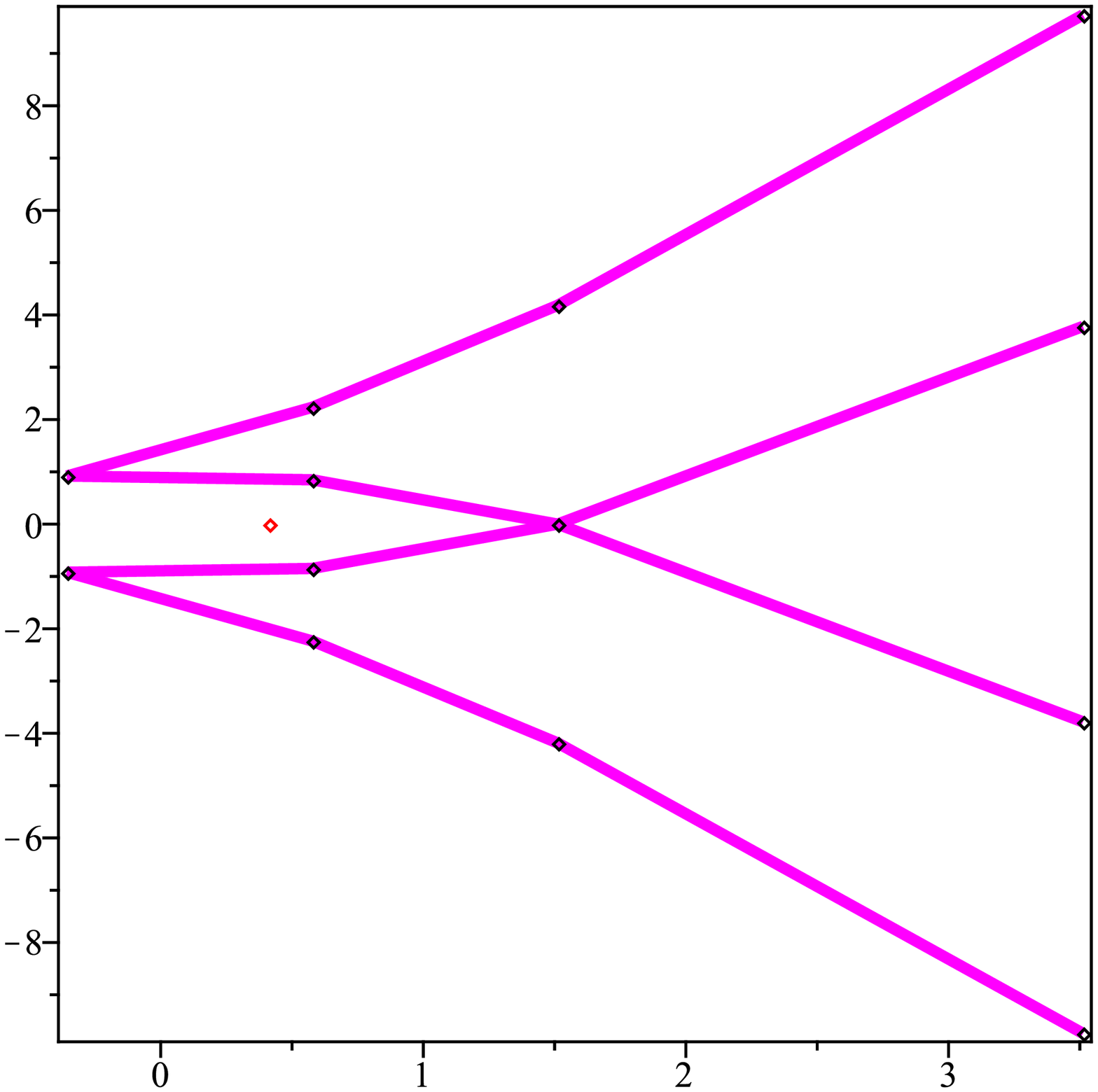,width=5cm,height=5cm}\\
\psfig{figure=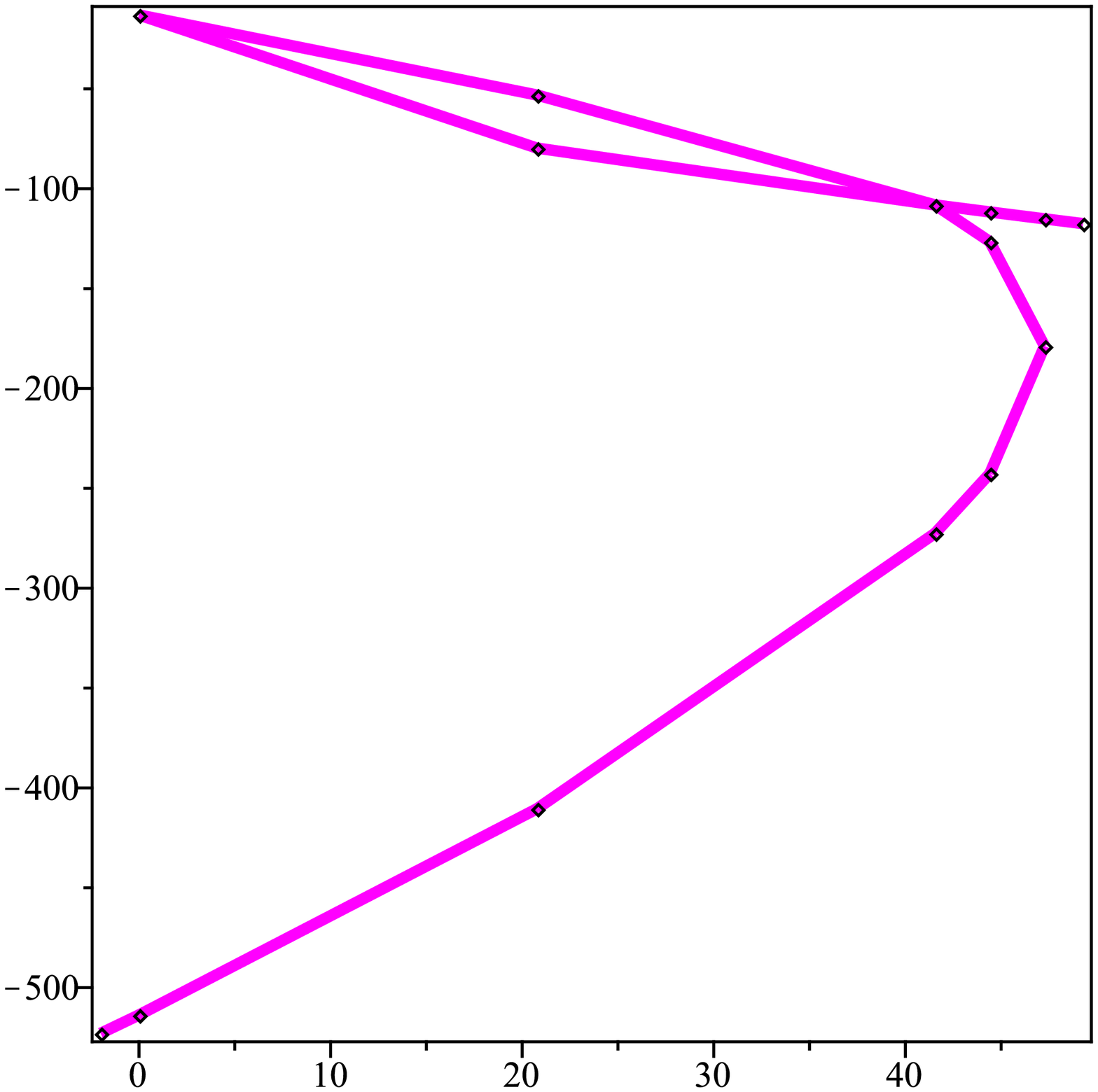,width=5cm,height=5cm} &
\psfig{figure=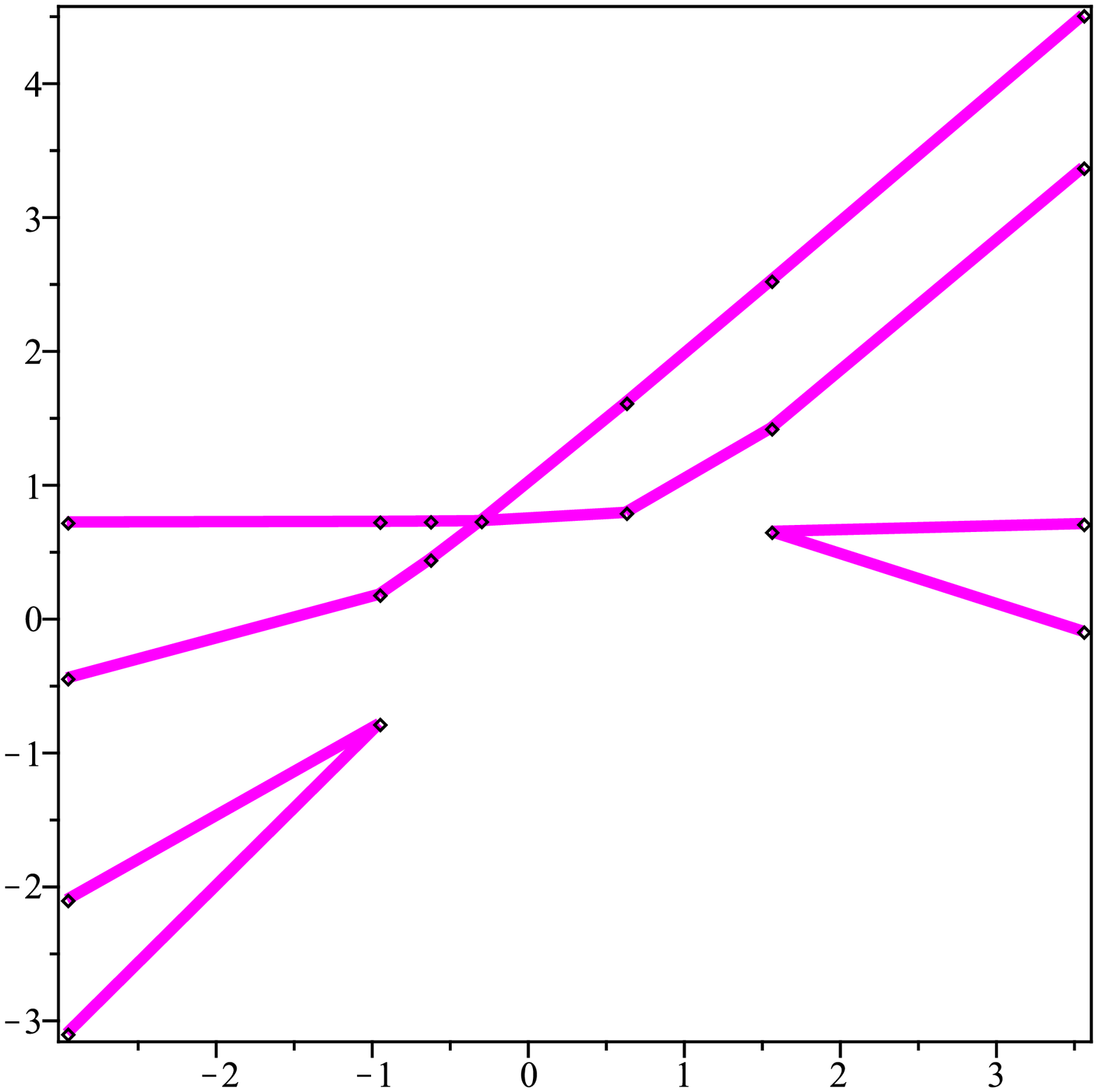,width=5cm,height=5cm} &
\psfig{figure=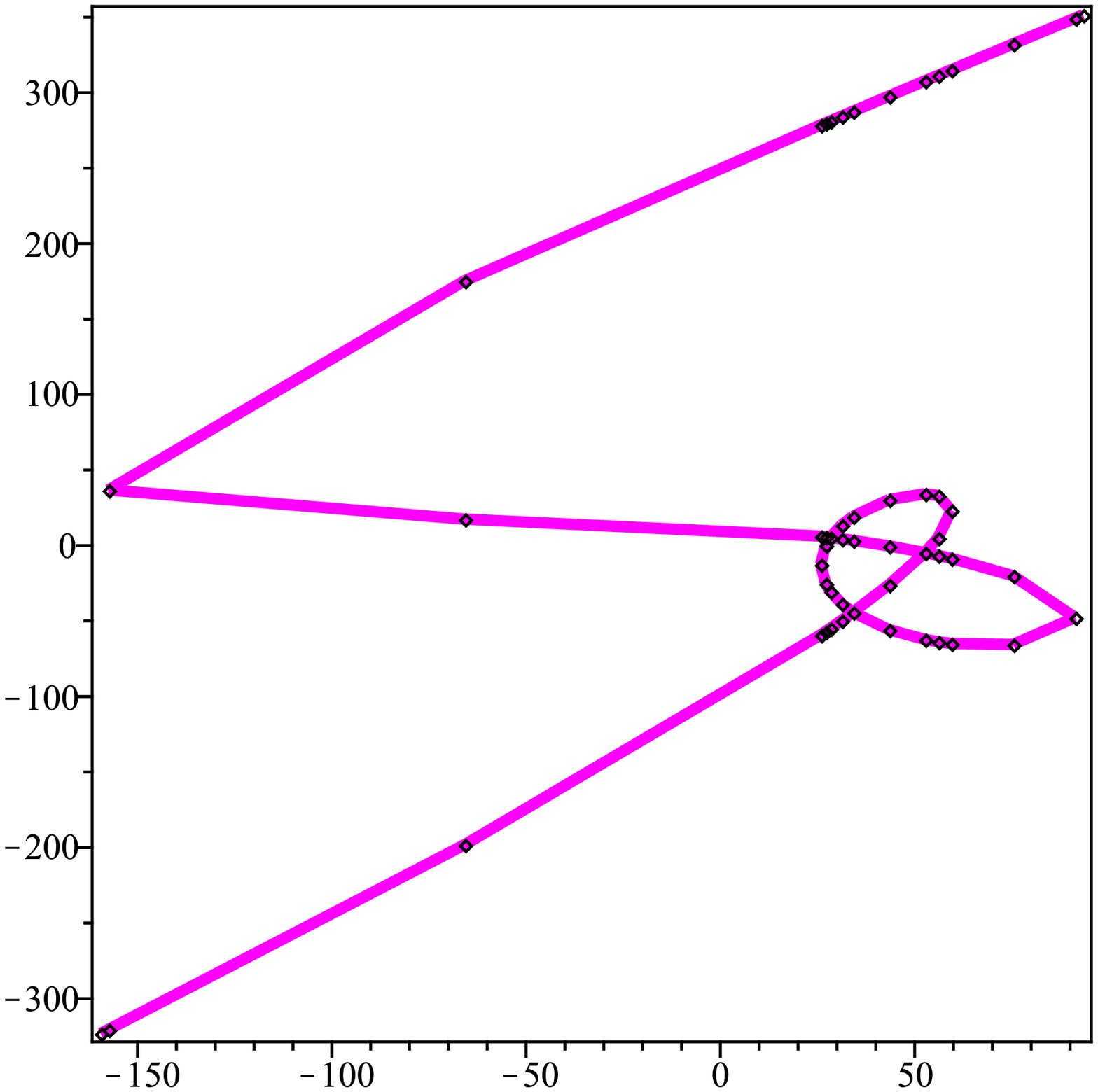,width=5cm,height=5cm}\\
\end{array}$}
\end{center}
\caption{Examples of the 2D algorithm.}
\end{figure}

Finally, we present examples of the 3D algorithm. In Table 2, for each curve we include: the degree of the
parametrization ($d_p$), the total degree of the implicit equation
of the projection onto the $xy$-plane ($d_i$), the number of terms
of this projection (n.terms), the timing without computing
isolated points ($t_0$), the timing including the
computation of isolated points ($t_1$), and the number of digits
used. As in the 2D-case, in all cases the computations start with
10 digits, and the algorithm increases the number of digits when
it is necessary. The parametrizations corresponding to these
examples are given in Appendix II.

\begin{center}
\begin{tabular}{|c|l|l|c|c|c|c|l|} \hline
Example & $d_p$ & $d_i$ & n.terms  & $t_0$ & $t_1$ & Digits  \\
\hline \hline 1 & 8 & 8 & 38  & 5.578 & 6.188 & 30  \\
\hline 2 & 10 & 10 & 65  & 3.516 & 3.297 & 10 \\
\hline 3 & 21 & 21 & 234  & 4.453 & 4.515 & 10 \\
\hline 4 & 4 & 7 & 8  & 0.657 & 0.625 & 10 \\
\hline 5 & 6 & 6 & 28  & 0.437 & 0.266 & 10 \\
\hline 6 & 8 & 4 & 5  & 0.141 & 0.109 & 10 \\
\hline 7 & 4 & 4 & 15  & 0.125 & 0.500 & 10 \\
\hline 8 & 12 & 12 & 91  & 1.015 & 0.875 & 10 \\
\hline 9 & 16 & 16 & 142  & 74.00 & 74.094 & 65\\
 \hline
\end{tabular}

{\bf Table 2:} 3D Examples.
\end{center}

The pictures corresponding to these curves can be found in Figure
3. The diamond in each picture points out the origin of the system
of coordinates; moreover, in Example 7 we have not plotted the
axes for the isolated point to be better appreciated.

\begin{figure}[ht]
\begin{center}
\centerline{$\begin{array}{ccc}
\psfig{figure=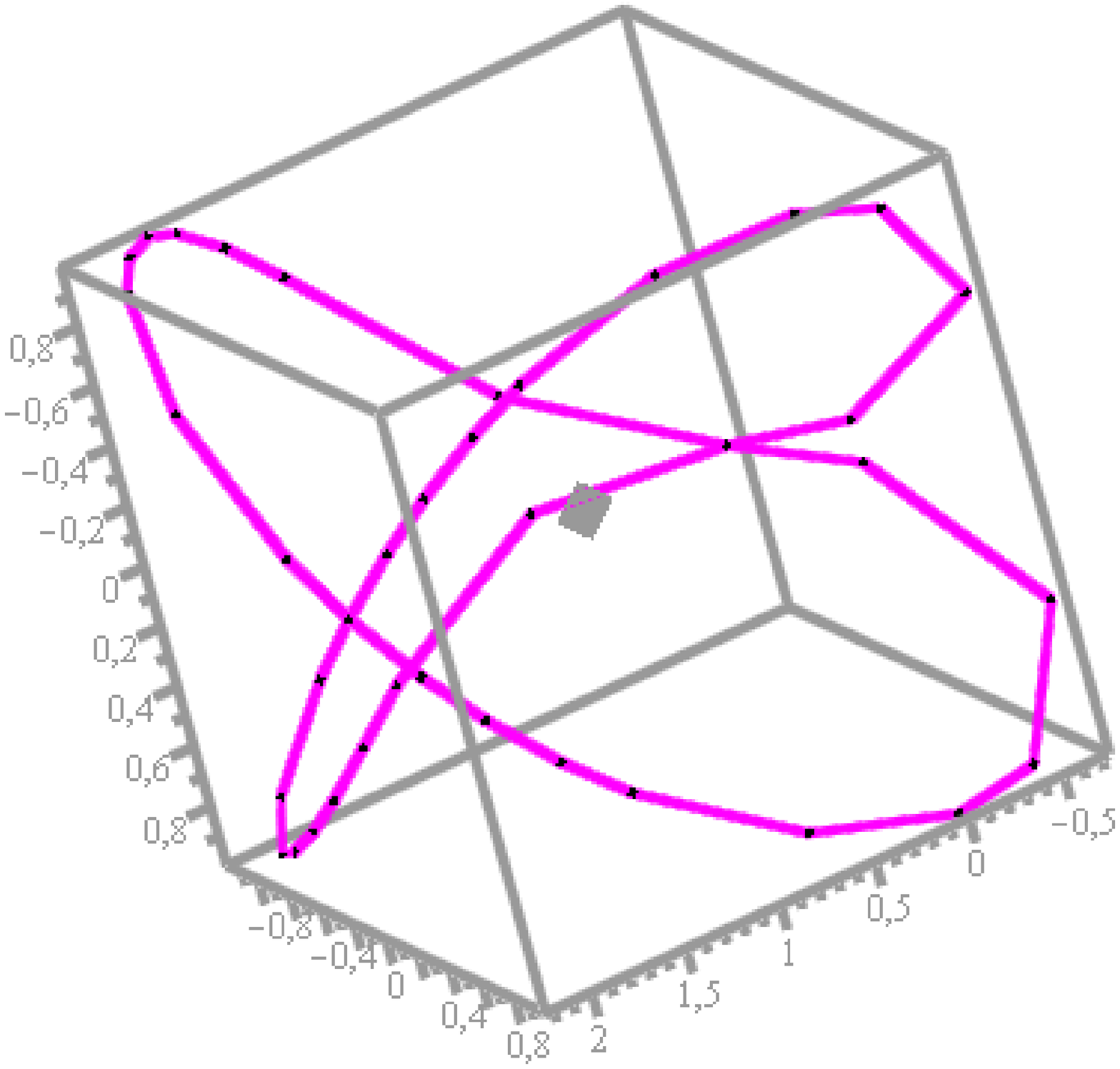,width=5cm,height=5cm} &
\psfig{figure=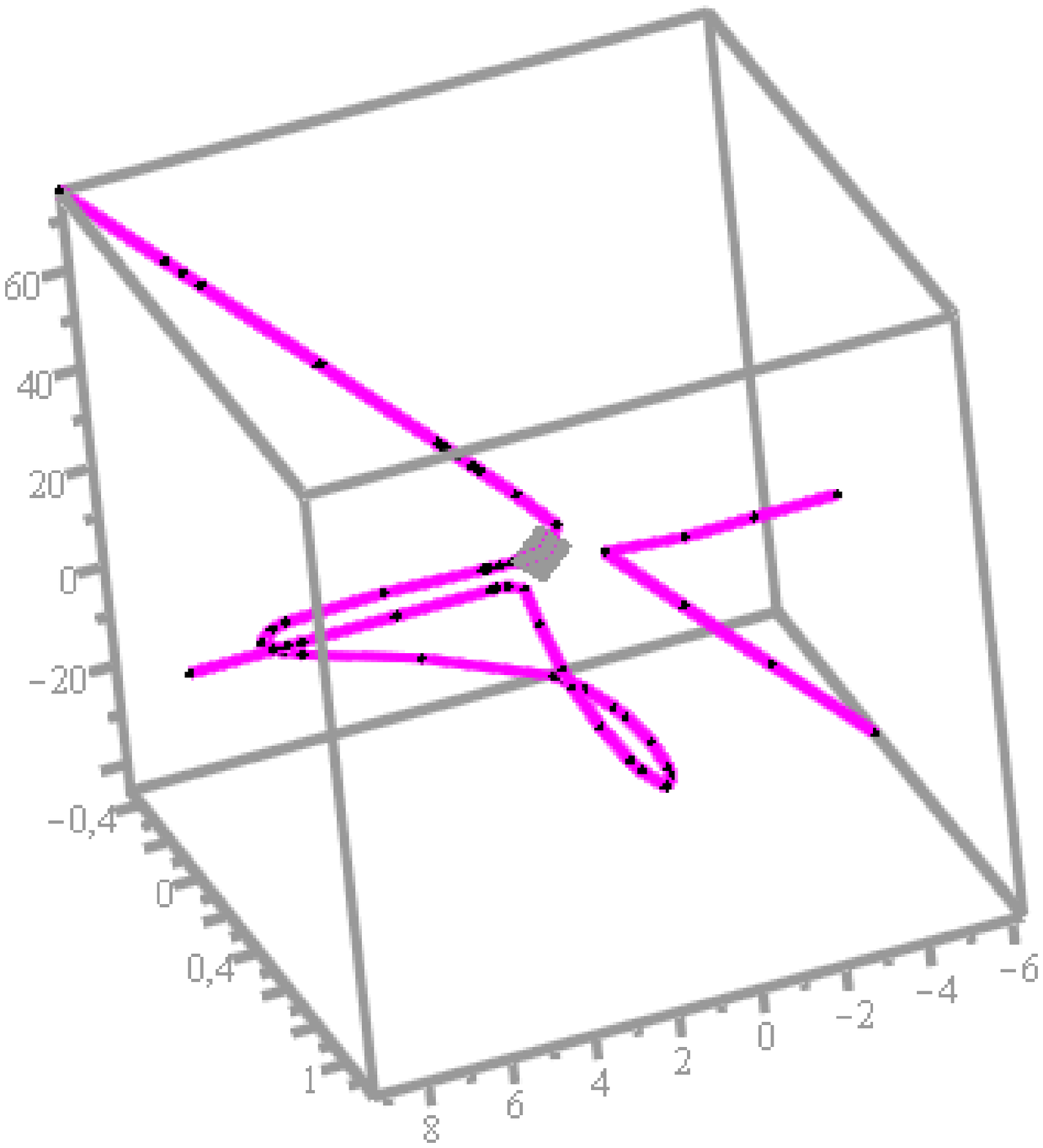,width=5cm,height=5cm} &
\psfig{figure=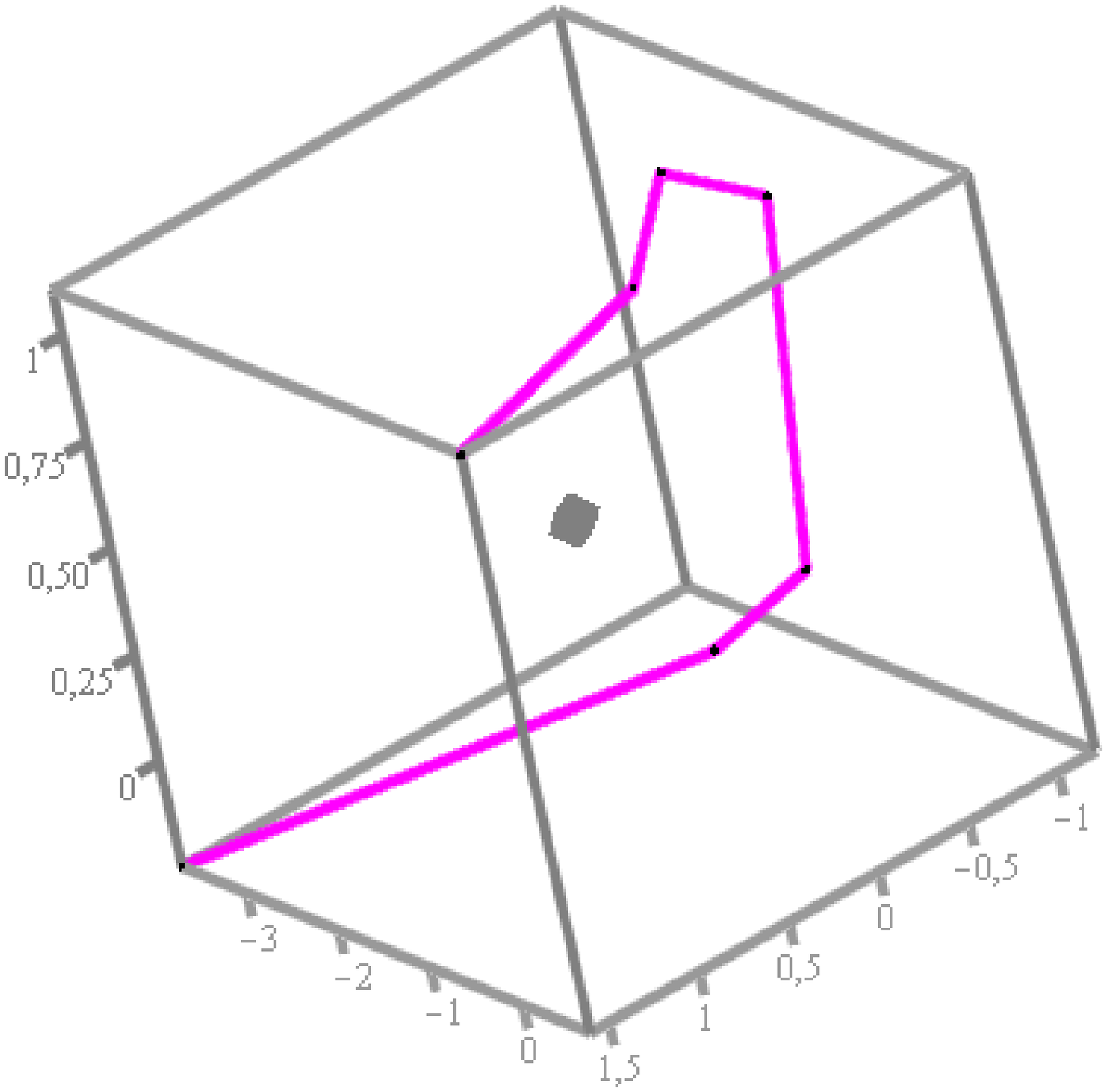,width=5cm,height=5cm} \\
\psfig{figure=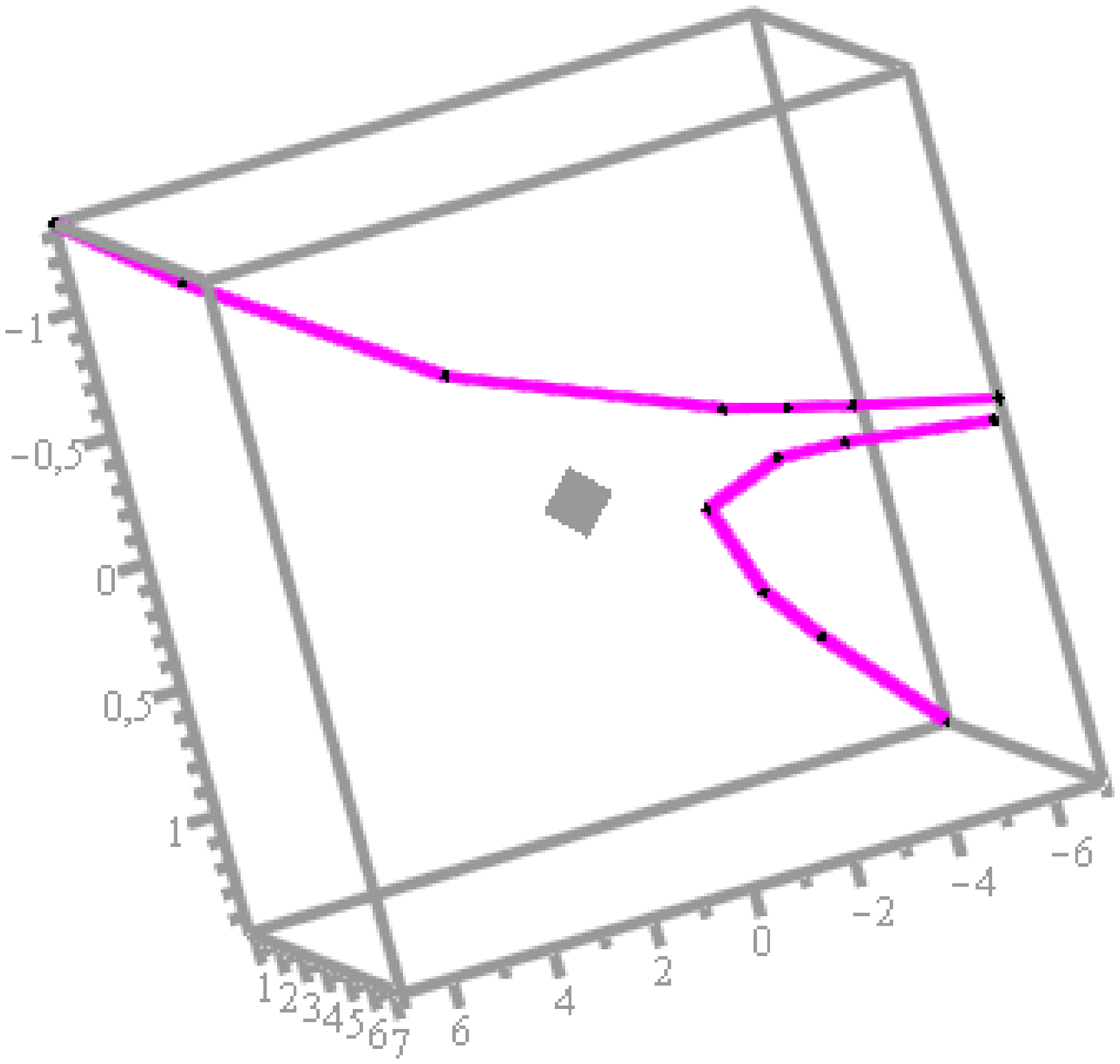,width=5cm,height=5cm} &
\psfig{figure=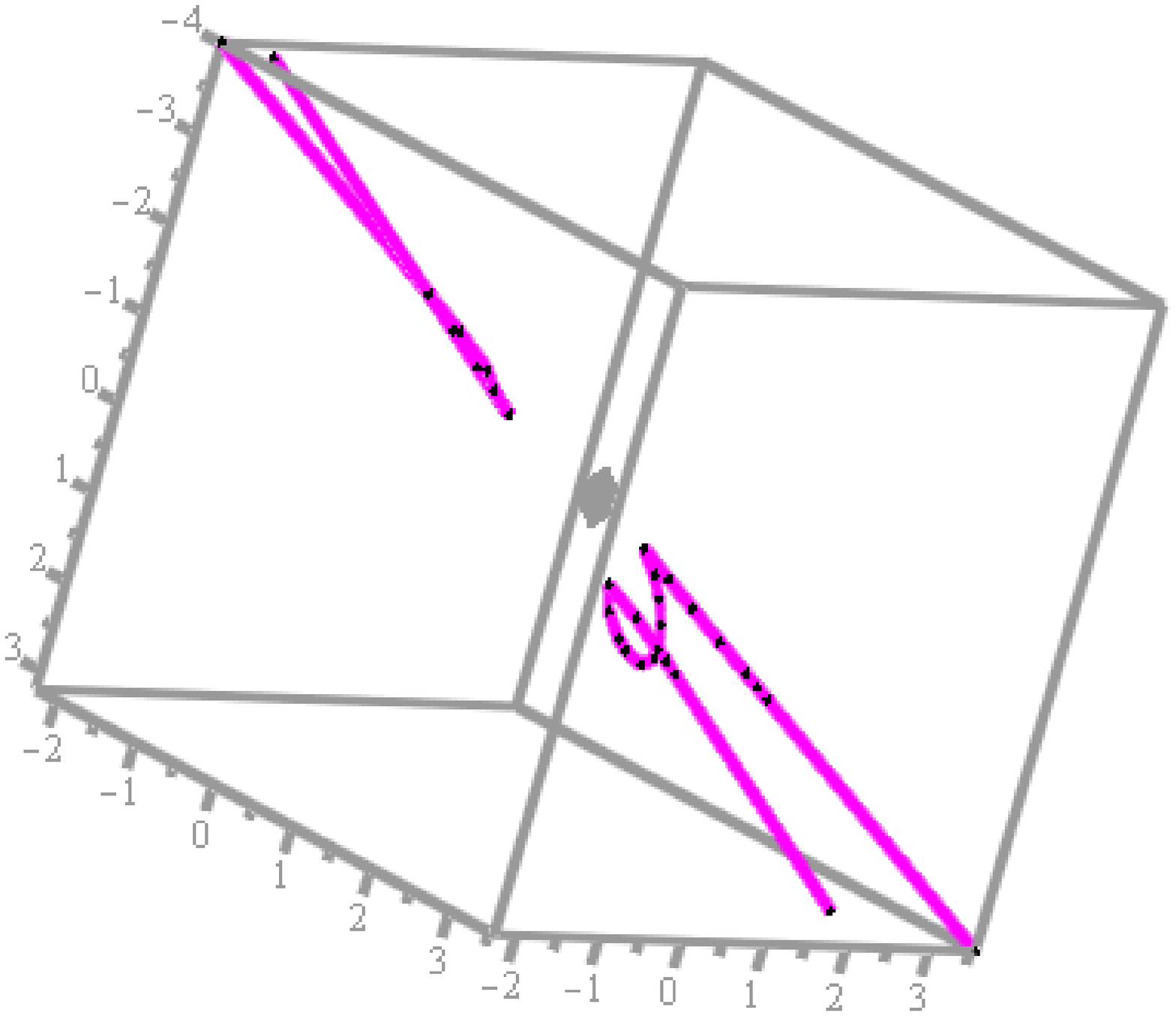,width=5cm,height=5cm} &
\psfig{figure=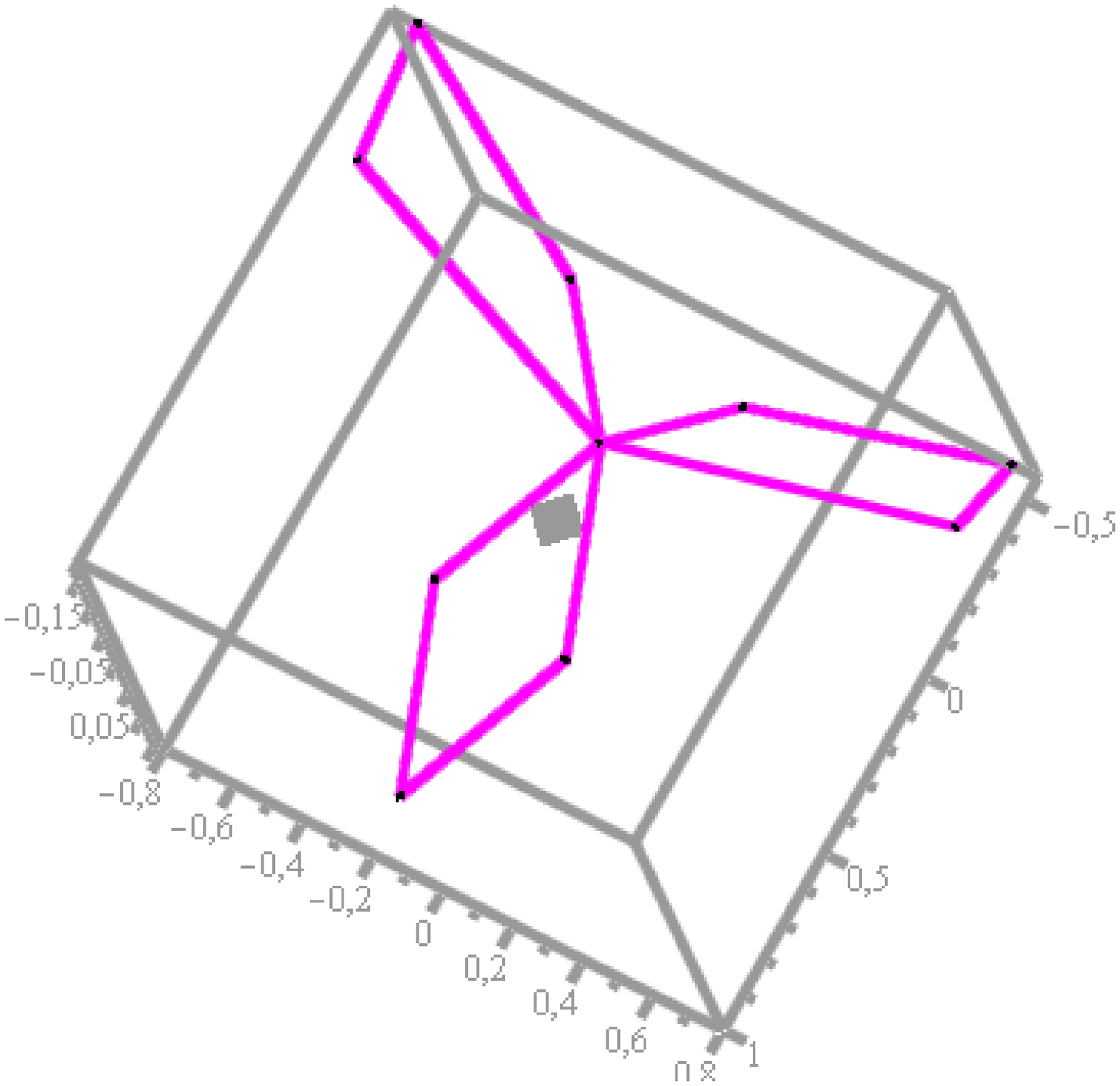,width=5cm,height=5cm}\\
\psfig{figure=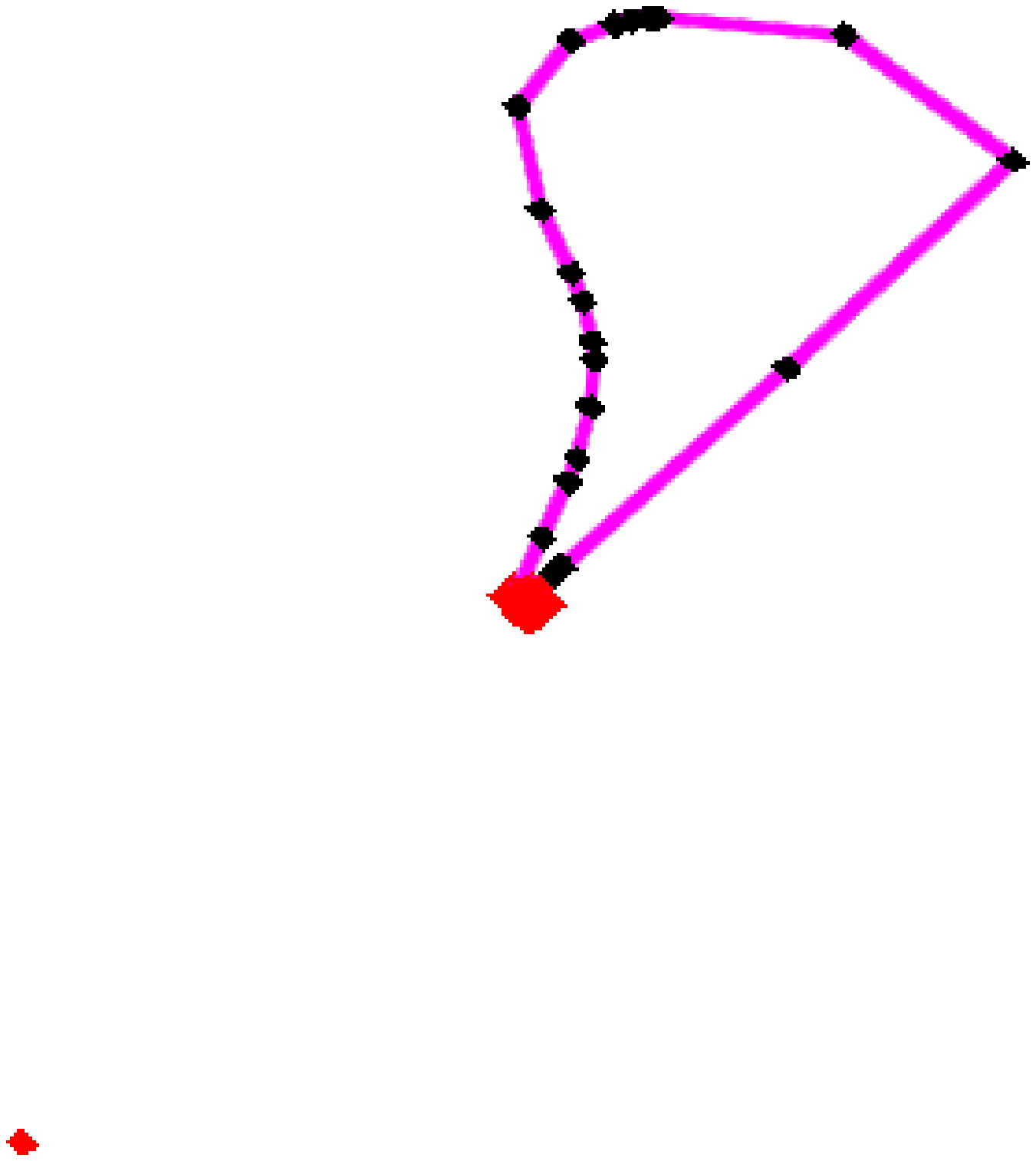,width=5cm,height=5cm} &
\psfig{figure=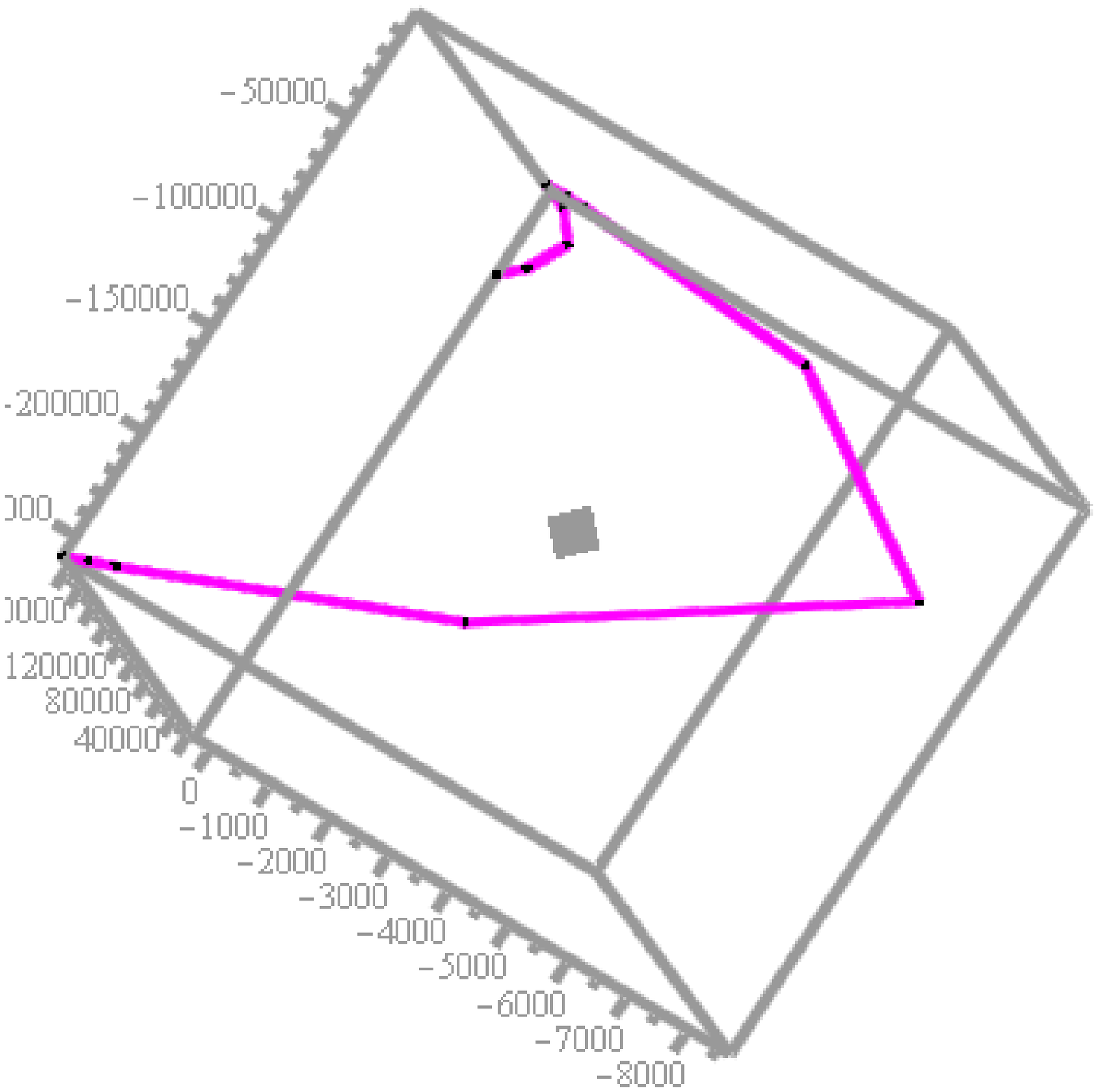,width=5cm,height=5cm} &
\psfig{figure=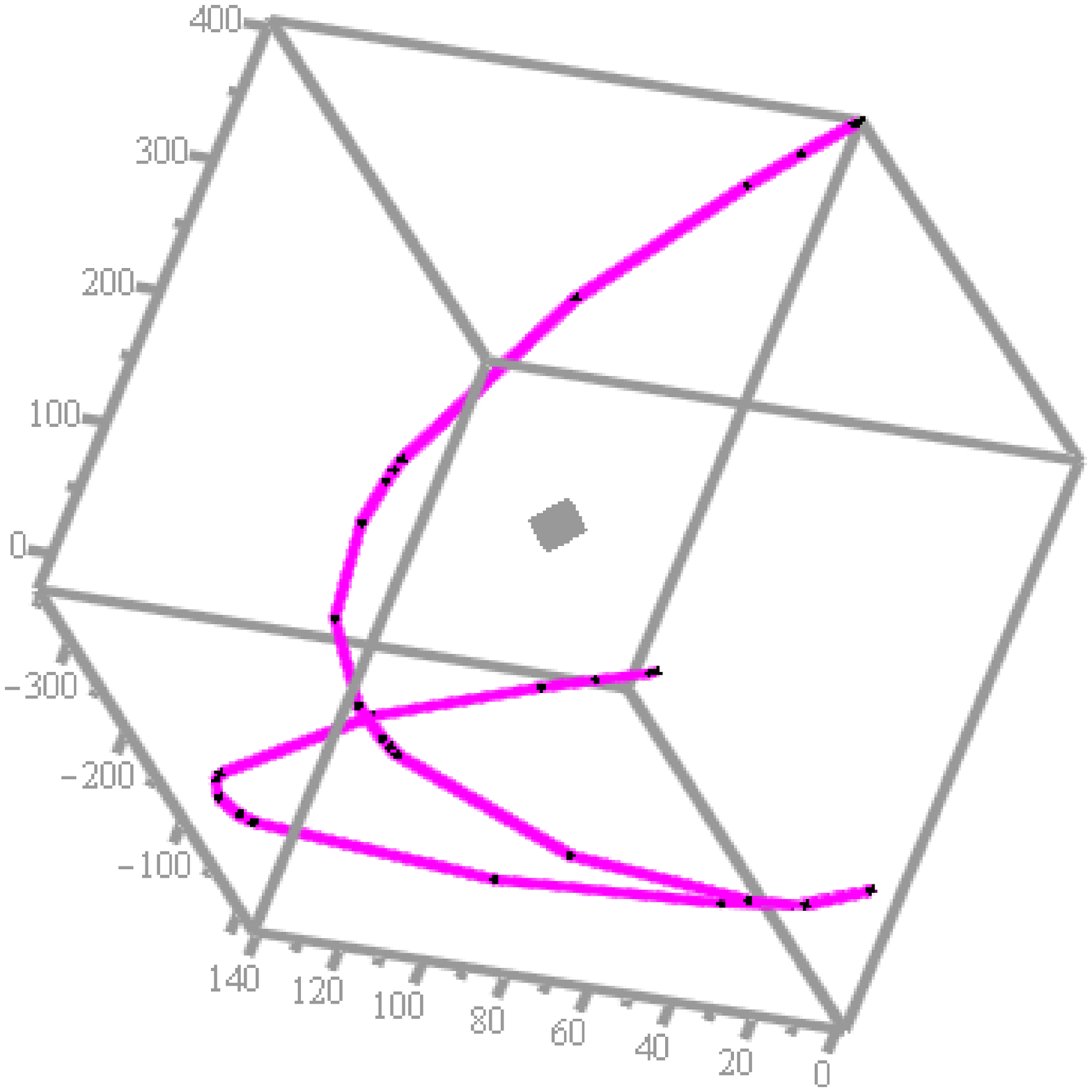,width=5cm,height=5cm}\\
\end{array}$}
\end{center}
\caption{Examples of the 3D algorithm.}
\end{figure}

\section{Acknowledgements}
We want to thank Prof. J.R. Sendra for his excellent ideas and the
time and energy he devoted to us. We also want to thank Prof.
Gonz\'alez-Vega for suggesting the problem and discussing it with
us.

\newpage

\section*{Appendix I: Parametrizations of the planar curves used in
the experimentation}\label{appen}

\underline{Example 1:}

$\varphi(t)=\left(\displaystyle{\frac{44+37t^3-23t^2+87t}{10+29t^3+98t^2-23t}},\mbox{
}
\displaystyle{\frac{95-61t^3-8t^2-29t}{40+11t^3-49t^2-47t}}\right)$

\underline{Example 2:}

$\varphi(t)=\left(\displaystyle{\frac{3456t^5-31104t^3+6t^8-756t^6+61236t^2-39366}{486t^4+36t^6+2916t^2+t^8+6561}},\right.
$\\
$\left.\displaystyle{\frac{-18t(864t^3-16t^5-1296t+6t^6-126t^4-1134t^2+4374)}{486t^4+36t^6+2916t^2+t^8+6561}}\right)$

\underline{Example 3:}

$\varphi(t)=\left(t^8-8t^6+20t^4-16t^2+2,t^7-7t^5+14t^3-7t\right)$

\underline{Example 4:}

$\varphi(t)=\left(\displaystyle{\frac{-7t^4+288t^2+256}{t^4+32t^2+256},\frac{-80t^3+256t}{t^4+32t^2+256}}\right)$

\underline{Example 5:}

$\varphi(t)=\left(\displaystyle{\frac{3t^2+3t+1}{-3t-1+t^6-2t^4},\frac{t^2(t^4-2t+2)}{-3t-1+t^6-2t^4}}\right)$

\underline{Example 6:}

$\varphi(t)=\left(\displaystyle{\frac{(t^2-1)(t^4-1+9t^2)}{9t^2(t^2+1)},\frac{-(t^8-2t^6+2t^2-1-54t^4)}{27(t^2+1)t^3}}\right)$

\underline{Example 7:}

$\varphi(t)=\left(-83t^{23}+98t^{20}-48t^{18}-19t^{13}+62t^{11}+37t^8,-13-64t^{27}+64t^{25}-90t^{22}-60t^{12}-34t^2\right)$

\underline{Example 8:}

$\varphi(t)=\left(\displaystyle{\frac{9+85t^6+80t^5+90t^3+74t^2+27t}{5-91t^6+81t^5+65t^4-12t^2+78t},
\frac{-56-5t^6+36t^5-8t^4+30t^3-3t}{-79-70t^5+42t^4+9t^3-21t^2-27t}}\right)$

\underline{Example 9:}

$\varphi(t)=\left(t^{17}+80-20t^5-4t^4-89t^3-77t^2+69t,
t^{17}-64-33t^6+21t^4-35t^3+97t^2+30t\right)$

\section*{Appendix II: Parametrizations of the space curves used in
the experimentation}\label{appen-2}

For each example, will use the notation
$\varphi(t)=(x(t),y(t),z(t))$.

\underline{Example 1:}

$x(t)=\displaystyle{\frac{36t(-1-98t-3954t^2-78868t^3-726692t^4-1092840t^5+31242296t^6+193263952t^7}{q(t)}}$

$y(t)=\displaystyle{\frac{-648t^2(1+84t+2940t^2+54880t^3+549996t^4+2492112t^5+2385712t^6)}{q(t)}}$

$z(t)=\displaystyle{\frac{36t(476t^3+426t^2+42t+1)}{64660t^4+10976t^3+1176t^2+56t+1}},$

with
$q(t)=112t+5488t^2+153664t^3+2741608t^4+33057472t^5+272552896t^6+1419416320t^7+4180915600t^8+1$

\underline{Example 2:}

$x(t)=\displaystyle{\frac{7+33t^{10}+80t^9-57t^7+88t^3+75t^2}{5t^{10}+61t^8+8t^7+71t^6-16t^5+37t}}$

$y(t)=\displaystyle{\frac{18t^8+28t^7+58t^5+69t^4+8t^3+4t}{5t^{10}+61t^8+8t^7+71t^6-16t^5+37t}}$

$z(t)=\displaystyle{\frac{-94t^9-59t^5+16t^4-82t^3+69t^2-t}{5t^{10}+61t^8+8t^7+71t^6-16t^5+37t}}$

\underline{Example 3:}

$\varphi(t)=\left(\displaystyle{\frac{t^{20}+t-1}{t^2+1}},\displaystyle{\frac{t^{21}-2}{t^2+1}},\displaystyle{\frac{t^5+1}{t^2+1}}\right)$

\underline{Example 4:}

$\varphi(t)=\left(\displaystyle{\frac{t^2+1}{t^4+1}},\displaystyle{\frac{1}{t^3}},t^2
\right)$

\underline{Example 5:}

$x(t)=\displaystyle{\frac{(t-1)^4(1+4t+7t^2)}{1-4t+17t^2-5t^6-13t^4+20t^5+48t^3}}$

$y(t)=\displaystyle{\frac{(1-4t+22t^2-4t^3+t^4)(1+t)^2}{1-4t+17t^2-5t^6-13t^4+20t^5+48t^3}}$

$z(t)=\displaystyle{\frac{(1-4t+22t^2-4t^3+t^4)(1+t)^2}{1-4t+17t^2-5t^6-13t^4+20t^5+48t^3}}$

\underline{Example 6:}

$\varphi(t)=\left(\displaystyle{\frac{1-3t^2}{(t^2+1)^2}},\displaystyle{\frac{(1-3t^2)t}{(t^2+1)^2}},\displaystyle{\frac{(1-3t^2)t^3}{(t^2+1)^4}}\right)$

\underline{Example 7:}

$x(t)=\displaystyle{\frac{87-7t^4+22t^3-55t^2-94t}{-73-56t^4-62t^2+97t}}$

$y(t)=\displaystyle{\frac{-82-4t^4-83t^3-10t^2+62t}{-73-56t^4-62t^2+97t}}$

$z(t)=\displaystyle{\frac{-82-4t^4-83t^3-10t^2+62t}{-73-56t^4-62t^2+97t}}$

\underline{Example 8:}

$x(t)=91+11t^{12}-49t^{10}-47t^7+40t^6-81t$

$y(t)=-28t^{12}+16t^{10}+30t^8-27t^5-15t^3-59t^2$

$z(t)=53+43t^{10}+92t^9-91t^6-88t^3-48t$

\underline{Example 9:}

$x(t)=-90t^{16}+81t^8+65t^6-12t^5+78t^4+5t^3$

$y(t)=-70t^{16}+42t^{15}+9t^{12}-21t^9-27t^8-79t^5$

$z(t)=62-14t^{14}+83t^{12}-96t^7-8t^3-54t^2$


\newpage

\begin{thebibliography}{56}

\bibitem{Arnon} Arnon
D.,  MacCallum S. (1988). {\it A polynomial time algorithm for the
topology type of a real algebraic curve}, Journal of Symbolic
Computation, vol. 5 pp 213-236.

\bibitem{JG-Sendra} Alcazar J.G., Sendra R. (2005) {\it Computation of
the Topology of Real Algebraic Space Curves}, Journal of Symbolic
Computation 39, pp. 719-744.



\bibitem{Tomas} Andradas C., Recio T., Sendra J.R. (1997) {\it A relatively optimal
reparametrization algorithm through canonical divisors}, Proceedings ISAAC 97, ACM press, pp. 349-355.

\bibitem{Andradas} Andradas C., Recio T. (2007) {\it Missing points and branches of real parametric curves}, Applicable Algebra in Engineering and Computing 18 (1-2), pp. 107-126


\bibitem{cox-1} Cox D., Little J., O'Shea D. (1992). {\it Ideals, Varieties and Algorithms}. Springer.

\bibitem{cox} Cox D., Little J., O'Shea D. (2005). {\it Using Algebraic Geometry}. Second Edition. Springer.

\bibitem{Diat} Diatta D.N., Mourrain B. and Ruatta O. (2008) {\it On the Computation of the Topology of a Non-Reduced Implicit Space Curve}. In Proceedings ISSAC 2008, ed. David Jeffrey, pp. 47-55.

\bibitem{Eigen} Eigenwilling A., Kerber M., Wolpert N. (2007) {\it Fast and Exact Geometric Analysis of Real Algebraic Plane Curves}, in C.W. Brown, editor, Proc. Int. Symp. Symbolic and Algebraic Computation, pp. 151-158, Waterloo, Canada. ACM.

\bibitem{ElKa} El Kahoui M. (2008) {\it Topology of Real Algebraic Space Curves}. Journal of Symbolic
Computation vol. 43, pp. 235-258.

\bibitem{gianni} Gianni P.,
Traverso C. (1983). {\it Shape determination of real curves and
surfaces}, Ann. Univ. Ferrera Sez VII Sec. Math. XXIX pp 87-109.

\bibitem{LaloCompl} Gonzalez-Vega L.,   El
Kahoui M. (1996). {\it  An improved upper complexity bound for the
topology computation of a real algebraic plane curve}, J.
Complexity 12 pp 527-544.

\bibitem{Lalo} Gonzalez-Vega L., Necula I. (2002).
{\it Efficient topology determination of implicitly defined
algebraic plane curves}, Computer Aided Geometric Design, vol. 19
pp. 719-743.

\bibitem{Hong} Hong H. (1996). {\it An effective
method for analyzing the topology of plane real algebraic curves},
Math. Comput. Simulation 42 pp. 571-582

\bibitem{Sonia} P\'erez-D\'{\i}az S. (2007). {\it Computation of the singularities of parametric plane curves}, Journal of Symbolic Computation, vol. 42, pp. 835-857.


\bibitem{Rubio} Rubio R., Serradilla J.M., V\'elez M.P. (2008). {\it Detecting real singularities of a space curve from a real rational parametrization}, Journal of Symbolic Computation, etc.



\bibitem{Seder} Sederberg T.W. (1986). {\it Improperly parametrized rational curves}, Computer Aided Geometric Design 3, 67-75.

\bibitem{seidel} Seidel R., Wolpert N. (2005) {\it On the Exact Computation of the
Topology of Real Algebraic Curves}. Proc. of the 21st Ann. ACM Symp. on Comp. Geom. (SCG 2005). ACM, 2005 107--115.

\bibitem{S02} Sendra J.\,R. (2002).
{\it Normal Parametrizations of Algebraic Plane Curves}.
 Journal of Symbolic Computation vol. 33, pp. 863--885.




\bibitem{SWPD} Sendra J.R., Winkler F., Perez-Diaz P. (2008). {\it Rational Algebraic Curves}, Springer-Verlag.



\end{thebibliography}
\end{document}